\documentclass{aa}

\usepackage[utf8]{inputenc}
\usepackage{color}
\usepackage{amsmath}
\usepackage{multirow}

\title{Comprehensive stellar seismic analysis}
\titlerunning{Comprehensive stellar seismic analysis}
\subtitle{New method exploiting the glitches information in solar-like pulsators}
\author{M.~Farnir\inst{\ref{ULg}}
\and M-A.~Dupret\inst{\ref{ULg}}
\and S.J.A.J.~Salmon\inst{\ref{ULg}}
\and A. Noels\inst{\ref{ULg}}
\and G. Buldgen\inst{\ref{Bir}}}
\institute{Institut d’Astrophysique et Géophysique de l’Université de Liège, Allée du 6 août 17, 4000 Liège, Belgium \\ \email{martin.farnir@uliege.be}\label{ULg}
\and School of Physics and Astronomy, University of Birmingham, Edgbaston, Birmingham B15 2TT, UK
\label{Bir}}
\date{Received <date> /
Accepted <date>}

\abstract {} {We develop a method that provides a comprehensive analysis of the oscillation spectra of solar-like pulsators. We define new seismic indicators that should be as uncorrelated and as precise as possible and should hold detailed information about stellar interiors. This is essential to improve the quality of the results obtained from asteroseismology as it will provide better stellar models which in turn can be used to refine inferences made in exoplanetology and galactic archaeology.} {The presented method -- WhoSGlAd -- relies on \emph{Gram-Schmidt}'s orthogonalisation process. A Euclidean vector subspace of functions is defined and the oscillation frequencies are projected over an orthonormal basis in a specific order. This allows the obtention of independent coefficients that we combine to define independent seismic indicators.} {The developed method has been shown to be stable and to converge efficiently for solar-like pulsators. Thus, detailed and precise inferences can be obtained on the mass, the age, the chemical composition and the undershooting in the interior of the studied stars. However, attention has to be paid when studying the helium glitch as there seems to be a degeneracy between the influence of the helium abundance and that of the heavy elements on the glitch amplitude. As an example, we analyse the 16CygA (\object{HD 186408}) oscillation spectrum to provide an illustration of the capabilities of the method.} {}

\begin{document}
\maketitle
\section{Introduction}
Since the launch of CoRoT \citep{2009IAUS..253...71B} and \emph{Kepler} \citep{2010AAS...21510101B} missions, the scientific community has access to a tremendous amount of asteroseismic data of unprecedented quality. Such data are essential to better constrain stellar structure and evolution and, in turn, improve the characterisation of exoplanets and stellar populations. However, it is essential to develop techniques that are able to retrieve stellar parameters as accurately as possible in order to benefit from the quality of the data.
A very complex problem in determining stellar parameters is the model dependency of the results. The results are intrinsically dependent on the input physics such as the equation of state as well as the opacity tables used. It therefore becomes of prime importance to develop techniques that are able to test the influence of the input physics on stellar parameters or even techniques that provide results as model independent as possible.

With such precision in the data, and the precision of the future missions TESS \citep{2014SPIE.9143E..20R} and PLATO \citep{2014ExA....38..249R}, studying the signature of acoustic glitches becomes a natural step towards better models. This idea was originally proposed by \citet{1988IAUS..123..151V} and \citet{1990LNP...367..283G}. They both highlighted the effect of a sharp feature in the stellar structure on the frequencies, either directly or on the second differences. Such considerations have already been the subject of several studies. As an example of the several techniques used, \citet{2014ApJ...782...18M} illustrate four techniques either using the second frequency differences or striving to isolate the glitch oscillation directly from the frequencies. Also, most of the current methods focus on the localisation of the helium second ionisation zone or the base of the envelope convection zone (e.g. \citealt{2000MNRAS.316..165M}). This is a crucial first step on the way to a better understanding of stellar physics. Indeed, a characterisation of the convective envelope extension allows to get constraints on convection itself as well as on overshooting (both its efficiency and nature). Therefore, the glitch provides the necessary observational data needed to refine current convection theories. Also, the study of the helium glitch should provide information on the helium surface content in low-mass stars. This is essential as, in such stars, it cannot be derived from spectroscopic data. Therefore, a methodology taking advantage of most of the oscillation spectrum aspects is required.

Finally, it is sometimes the case that studies use correlated constraints or discard pieces of information in their studies. For example, in method C from \citet{2014ApJ...790..138V}, the information that is not contained in the glitch (the smooth component) is not used directly as the glitch information is isolated to draw inferences. Therefore, the usual indicators (e.g. large separation, small separation ratios,...) are computed separately from the glitch and no method is proposed to determine properly the correlation between those indicators and that of the glitch.

For those reasons, we propose a method: \emph{WhoSGlAd} -- for \textbf{Who}le \textbf{S}pectrum and \textbf{Gl}itches \textbf{Ad}justment -- that takes as much of the available spectral information as possible into account. This method defines new seismic indicators in such a way that they are as independent as possible and significant from a statistical point of view. To do so, it relies on linear algebra via \emph{Gram-Schmidt}'s algorithm.

The present paper is organised as follows. We first present the method in a very general and mathematical way in Sect.~\ref{Sec:Met}. Then Sect.~\ref{Sec:SeiInd} defines the seismic indicators and their diagnosis power will be used to study solar-like pulsators. In the following section, we demonstrate its ability to extract and analyse the glitches signal. We also show its limitations. We present a first application to the case of 16 Cygni A (\object{HD 186408}) in Sect.~\ref{Sec:16Cyg}. Let us insist on the fact that we do not present here a thorough study of 16 Cygni A but we rather show an example of the ability of the method to provide constraints on stellar structure. We conclude the paper by discussing the results and detailing future perspectives.

\section{Method}\label{Sec:Met}
In the present section, we describe the method we developed. It aims at using as much as possible of the information available in the oscillation spectrum of a star. Therefore, both the oscillatory and smooth part of the spectrum are simultaneously analysed in a single adjustment. This avoids multiple usage of the same information to draw different inferences. The very strength of the proposed method is that the different parameters obtained will be independent of each other, i.e. their covariance matrix will be the identity matrix. This will allow to build indicators which also are independent of each other and draw statistically relevant inferences. The independence of the parameters will be ensured by using \emph{Gram-Schmidt}'s \citep{Gram,Schmidt} algorithm. Then, the defined seismic indicators will be used as constraints to provide improved models in the framework of forward seismic modelling (see for example \citet{2005A&A...441..615M} for one of the first use of \emph{Levenberg-Marquardt}'s algorithm to adjust a model to seismic and non-seismic observables). Finally, such models may be used as initial guesses for inverse seismic modelling. (see \citet{2002ESASP.485..337R,2002ESASP.485..341R} for the application of the inversion technique on an artificial target, which shows the feasibility of such techniques, and \citet{2016A&A...585A.109B,2016A&A...596A..73B} for examples of inversions in the case of 16 Cygni A.)

\subsection{Mathematical description}
Non-radial pulsation frequencies can be mathematically defined by three integer numbers; the radial order $n$, the spherical degree $l$, and the azimuthal order $m$ (in this paper, we do not consider the seismic probing of rotation and consider only $m=0$).
The method we developed -- \emph{WhoSGlAd} -- is based on linear algebra in a Euclidean space.
The vector space we consider is the set of $N$ observed oscillation frequencies $\nu_i$. The standard deviation for each frequency is written $\sigma_i$.
Given two frequency vectors $\mathbf{x}$ and $\mathbf{y}$ we define their scalar product as:
\begin{equation}\label{Eq:ScaPro}
\left\langle \mathbf{x} | \mathbf{y} \right\rangle~=~\sum\limits^N_{i=1} \frac{x_i y_i}{\sigma^2_i}.
\end{equation}

Often in asteroseismology, it is useful to compare two sets of frequencies (e.g. theoretical and observed frequencies) using a merit function defined as: 
\begin{equation}\label{Eq:Chi}
\chi^2~=~\sum\limits^N_{i=1} \frac{\left(\nu_{\textrm{obs},i}-\nu_{\textrm{th},i}\right)^2}{\sigma^2_i},
\end{equation} 
with $\boldsymbol{\nu}_\textrm{th}$ and $\boldsymbol{\nu}_\textrm{obs}$, the theoretical and observed\footnote{We denote by the subscript \textit{obs} both the observed frequencies and the frequencies derived from a reference model -- which constitutes an artificial target -- and we denote by the subscript \textit{th} the frequencies that we adjust to those observations.} frequencies.
Taking advantage of the scalar product defined above and the associated norm, this simply becomes: 
\begin{equation}\label{Eq:ChiNor}
\chi^2~=~\Vert\boldsymbol{\nu}_\textrm{obs}-\boldsymbol{\nu}_\textrm{th}\Vert^2.
\end{equation} 
In the presence of a glitch, \citet{2007MNRAS.375..861H} showed that the oscillatory component in frequencies due to the glitch can be isolated from the rest of the spectrum, called the smooth component. Thus, to represent observed frequencies, we define a vector subspace that is typically a polynomial space -- the smooth component -- associated with an oscillating component -- the glitch --. The analytical formulation of those two components will be given in the following sections. This is very similar to what has been done by \citet{2014ApJ...790..138V}. 

The method consists in the projection of the observed and theoretical frequencies over the vector subspace. Then, we define seismic indicators from the projections. Their definitions are given in Sect.~\ref{Sec:SeiInd}.
To do so, it is useful to define an orthonormal basis over the vector subspace. This is done via \emph{Gram-Schmidt}'s orthogonalisation process associated with the definition of the scalar product (\ref{Eq:ScaPro}). For more information about this process as well as its equivalent form as a QR decomposition, the interested reader may read App. \ref{Ap:MatMet}.

If we write $j$ and $j_0$ the indices associated with the basis elements, $\mathbf{p}_{j}$ the former basis elements, $\mathbf{q}_{j_0}$ the orthonormal basis elements, and $R^{-1}_{j,j_0}$ the transformation matrix, we have:
\begin{equation}\label{Eq:Bas}
q_{j_0}(n,l)~=~\sum\limits_{j\leq j_0}R^{-1}_{j,j_0}p_{j}(n,l),
\end{equation} 
where the dependence in $n$ and $l$ translates that the basis elements are evaluated at each observed value of the radial order $n$ and the spherical degree $l$.

It is essential to note that the projections will be done in a specific order to obtain the lowest possible value of the merit function. This will be the subject of the following subsection.
Finally, we write $a_{j}~=~ \left\langle \boldsymbol{\nu} \vert \mathbf{q}_{j} \right\rangle$ the projections of the frequencies over the basis elements. The fitted frequencies will then be given by:
\begin{equation}\label{Eq:FreFit}
\nu_{f}(n,l)~=~\sum\limits_j a_{j} q_{j}(n,l).
\end{equation}
Let us add that, thanks to the orthonormalisation, the standard deviations of the coefficients $a_{j}$ are $\sigma\left(a_{j}\right)~=~1$ and they are independent (their covariance matrix is the identity).

\subsection{Smooth component}\label{Sec:Smo}
Now that the mathematical context is given, we may detail the vector subspace we selected to fit the smooth component of the frequency spectrum.
As the set of observed radial orders and $\sigma_i$ are usually different for each spherical degree and the smooth component depends on $l$, the smooth component basis elements depend on $l$. For each value of $l$, the frequencies will be projected over the different powers considered. We also point out that, for each spherical degree, the method requires at least the same number of frequencies as of powers considered in the formulation. The polynomials are then of the general form:
\begin{equation}\label{Eq:PolSmo}
p_{lk}(n,l')~=~\delta_{ll'} p_{k}(n),
\end{equation}
where $\delta_{ll'}$ is the \emph{Kronecker} delta which compares two spherical degrees $l$ and $l'$ , $p_{k}(n)$ is a polynomial in the radial order $n$ and $k$ represents its ordering. We note that the previously defined $j$ is now separated into two indices, the spherical degree $l$  and the ordered power $k$. For a better understanding, the spherical degree and ordering will be explicitly written for the smooth component transformation matrix as $R^{-1}_{l,k,k_0}$.
And the orthonormal basis elements are:
\begin{equation}\label{Eq:bSmo}
q_{lk}(n,l')~=~\delta_{ll'} q_{lk}(n),
\end{equation}
which yield:
\begin{equation}\label{Eq:aSmo}
a_{lk}~=~ \left\langle \boldsymbol{\nu} \vert \mathbf{q}_{lk} \right\rangle~=~\sum\limits_n \frac{\nu(n,l)q_{lk}(n,l)}{\sigma^2(n,l)},
\end{equation}
where thanks to the introduction of $\delta_{ll'}$ in Eqs. \ref{Eq:PolSmo} and \ref{Eq:bSmo}, the sum over $l'$ collapses over a fixed degree l. For the smooth component, we treat separately each spherical degree, the parameters associated to a given degree only depend on the frequencies of this degree.

According to the asymptotic theory of non-radial oscillations \citep{1986HiA.....7..283G}, we have the following formulation of the expected frequencies as a function of $n$ and $l$:
\begin{equation}\label{Eq:AsyFre}
\nu\left(n,l\right)~\simeq~\left( n+\frac{l}{2}+\epsilon \right)\Delta,
\end{equation}
where $\Delta~=~\left(2\int^{R_*}_0 \frac{dr}{c(r)} \right)^{-1}$ is the asymptotic large frequency separation, $c(r)$ is the adiabatic sound speed, and $R_*$ is the radius at the photosphere of the star.

It follows that the first two polynomials in $\mathbf{n}$ (taken from the right hand side of Eq. \ref{Eq:PolSmo}) used to depict the spectrum smooth component will be :
\begin{equation}
p_{0}(n)~=~1,
\end{equation}
\begin{equation}
p_{1}(n)~=~n.
\end{equation}
Then, to provide the best fit to the observed frequencies, we methodically tested several combinations of powers to find the set giving the best agreement with the observations -- the observations actually referring to theoretical models taken as observed stars in a set of calibrations --, hence the lowest $\chi^2$ value. We get:
\begin{equation}
p_{2}(n)~=~n^2.
\end{equation}

At this point, it is of prime importance to note that the construction of the basis via Gram-Schmidt's process will have to be done following the ordering of the degrees because it will allow us to associate the seismic indicators to the projection of the frequencies on the successive basis elements. For example, the projection of the frequencies on the $0$ order polynomial corresponds to a fit to a constant value. This estimates the mean frequency value.
Moreover, we did not include other degrees as the fit of the smooth component was already very good\footnote{App. \ref{Ap:NEx} shows that adding new elements to the set of basis functions is indeed not relevant.}. Also, adding higher order polynomials to the smooth component might account for some of the glitch oscillating features. It is essential to avoid such a behaviour as the definitions of the seismic indicators, and the inferences we draw from them, will be impacted.

Furthermore, we could also include a regularisation parameter $\lambda$ (as in \citet{2014ApJ...790..138V}, method C) in order to prevent those behaviours. This requires a new definition of the vector subspaces. The vectors are now of dimension $2N$. The first $N$ components are defined as before while the components from $N+1$ to $2N$ are $0$ for both the frequencies and the glitch polynomial, and the second derivative of the polynomial for the smooth part. The vectors are now:
\begin{eqnarray}
\boldsymbol{\nu} & \rightarrow & \boldsymbol{\nu'}~\hspace{.9em}=~\left(\nu_{i,i\leq N},0_{N<i\leq 2N} \right)\nonumber \\
\text{smooth} & \rightarrow & \boldsymbol{q}_{j}'~=~\left(\boldsymbol{q}_{j}, \frac{\partial^2 \boldsymbol{q}_{j}}{\partial n^2} \right) \nonumber \\
\text{glitch} & \rightarrow & \boldsymbol{q}_{j}'~=~\left(\boldsymbol{q}_{j},\boldsymbol{0} \right).
\end{eqnarray} 
Then, we have to define the scalar product of $\boldsymbol{x}$ and $\boldsymbol{y}$ as:
\begin{equation}\label{Eq:ScaSmo}
\left\langle \boldsymbol{x} | \boldsymbol{y} \right\rangle~=~\sum\limits^N_{i=1} \frac{x_i y_i}{\sigma^2_i}~+~\lambda^2 \sum\limits^{2N}_{i=N+1} x_i y_i.
\end{equation}
Therefore, using the definition (\ref{Eq:ChiNor}) with the new scalar product gives another value of the merit function.
The inclusion of the regularisation parameter allows to minimise the oscillation of the smooth component as the minimisation of the merit function will lead to a minimisation of the second derivatives of the smooth component. Let us note that, for the regularisation constant to have an influence on the results, it must at least be on the order of the inverse of the frequencies standard deviation.
However, we performed some tests with and without regularisation terms and it appears that the method is very stable without it. Moreover, we observed in many cases that the results were degraded when including it (See also Sect.~\ref{Ap:Lam} for an illustrative example). Therefore, it is not necessary to include these regularisation terms to properly extract the glitch in our method. As a consequence, the results presented in this paper do not include such terms. The fact that we use fewer fitting parameters than in \citet{2014ApJ...790..138V} (they consider polynomials up to the fourth degree whereas we only reach the second) might explain that the regularisation constant is not necessary in our case.
In addition, we note that using $\lambda~=~0$ leads to the classical $\chi^2$ fitting.

Finally, as we have three polynomials for each value of $l$, we built, for the smooth component, a vector subspace of dimension $\mathbf{3 \times l}$ which equals $\mathbf{12}$ if we have four values for $l$ (e.g. $0,1,2,3$). As hinted earlier, more than $\mathbf{3}$ observed frequencies of each degree are necessary to apply the developed method.

\subsection{Glitch}\label{Sec:MetGli}
The formulations used by \citet{2014ApJ...790..138V} and \citet{2007MNRAS.375..861H} allow us to fit properly the helium and convection zone glitches but they are highly non linear with respect to the free parameters. Below is the expression from \citet{2014ApJ...790..138V}, which we adapted:
\begin{eqnarray}\label{Eq:NuVer}
\delta\nu_{g,\textrm{Verma}} & = & \mathcal{A}_\textrm{He} \nu e^{-c_2 \nu^2} \sin\left(4\pi \tau_\textrm{He} \nu + \phi_\textrm{He} \right) \nonumber \\
      & + & \frac{\mathcal{A}_\textrm{CZ}}{\nu^2} \sin\left(4\pi \tau_\textrm{CZ} \nu + \phi_\textrm{CZ} \right),
\end{eqnarray}
where the first term takes the helium glitch into account and the second, the convection zone glitch. The quantities $\tau_\textrm{He}$, $\phi_\textrm{He}$, and $\mathcal{A}_\textrm{He}$ represent respectively the acoustic depth of the second ionisation zone of helium, the phase of the helium glitch, and its amplitude. The same goes for the quantities $\tau_\textrm{CZ}$, $\phi_\textrm{CZ}$, and $\mathcal{A}_\textrm{CZ}$ in the case of the base of the envelope convective zone. Finally, $c_2$ is the rate of decrease in amplitude of the helium glitch with the squared frequency.

Moreover, Eq.~(\ref{Eq:NuVer}) is implicit since the frequency appears on the right-hand side. To adjust at best the frequencies, it is therefore necessary to use non-linear least square fitting algorithms (e.g. \emph{Levenberg-Marquardt}'s method, genetic algorithms,...) which can be unstable and are very sensitive to the initial guesses on the optimal parameters. 

For this reason, we decided to adopt the following \textit{linearised} functions, expressed as a function of $\widetilde{n}~=~\left(n+\frac{l}{2}\right)$ for the helium glitch:
\begin{equation}
p_{\textrm{He}Ck}(\widetilde{n})~=~\cos\left( 4\pi T_\textrm{He}\widetilde{n}\right) \widetilde{n}^{-k},
\end{equation}
\begin{equation}
p_{\textrm{He}Sk}(\widetilde{n})~=~\sin\left( 4\pi T_\textrm{He}\widetilde{n}\right) \widetilde{n}^{-k},
\end{equation}
with $k~=~(4,5)$; and, for the convection zone glitch:
\begin{equation}
p_{\textrm{C}C}(\widetilde{n})~=~\cos\left( 4\pi T_\textrm{CZ}\widetilde{n}\right) \widetilde{n}^{-2},
\end{equation}
\begin{equation}
p_{\textrm{C}S}(\widetilde{n})~=~\sin\left( 4\pi T_\textrm{CZ}\widetilde{n}\right) \widetilde{n}^{-2}.
\end{equation}
To obtain the above expressions we replaced the value of the frequency $\nu$ by its first order approximation from the asymptotic formulation (\ref{Eq:AsyFre}): $\widetilde{n}\Delta$. We also defined $T_\textrm{He}~=~\tau_\textrm{He} \Delta$ and $T_\textrm{CZ}~=~\tau_\textrm{CZ} \Delta$.

Moreover, we approximated the exponential decrease in frequency by the combination of two polynomials in $\widetilde{n}\Delta$. The degrees $-4$ and $-5$ have been chosen to reproduce at best the decrease of the glitch amplitude towards high frequencies (which was described using a gaussian by \citealt{2007MNRAS.375..861H}). To do so, we compared the polynomial formulation with the exponential one (Eq.~\ref{Eq:NuVer}). Let us add that we adjusted the helium glitch using several sets of degrees and the method proved to be very stable and the results remained good. Both the fitted coefficients and defined indicators values were quasi unaffected by the choice of degrees.
The glitch then writes:
\begin{equation}
\delta\nu_{\textrm{He}}(\widetilde{n}) = \sum\limits^{4}_{k=5}\left[c_{\textrm{He},k} p_{\textrm{He}Ck}(\widetilde{n}) + s_{\textrm{He},k} p_{\textrm{He}Sk}(\widetilde{n}) \right],
\end{equation}
\begin{equation}
\delta\nu_{\textrm{CZ}}(\widetilde{n}) = \left[c_\textrm{CZ} p_{\textrm{C}C}(\widetilde{n}) + s_\textrm{CZ} p_{\textrm{C}S}(\widetilde{n}) \right].
\end{equation}

Furthermore, such a formulation allows us to move on with \emph{Gram-Schmidt}'s process and generate orthonormal vectors to append the smooth component basis. As the glitch is generated in the superficial layers, it should not depend on $l$. Therefore, we defined the coefficients $c$ and $s$ to be independent of $l$. By doing so, the vector subspace associated with the glitch is only of dimension $6$ and this subspace is used to complete the orthonormal basis over which frequencies are projected.
We are thus able to write the glitch contribution to the frequencies as:
\begin{eqnarray}
\delta\nu_{g}(\widetilde{n}) & = & \sum\limits^{4}_{k=5}\left[C_{\textrm{He},k}q_{\textrm{He}Ck}\left(\widetilde{n},T_\textrm{He}\right) +  S_{\textrm{He},k}q_{\textrm{He}Sk}\left(\widetilde{n},T_\textrm{He}\right)\right] \nonumber \\
      & + & C_\textrm{CZ}q_{\textrm{C}C}\left(\widetilde{n},T_\textrm{CZ}\right)+S_\textrm{CZ}q_{\textrm{C}S}\left(\widetilde{n},T_\textrm{CZ}\right).
\end{eqnarray}
We draw the reader's attention to the fact that the basis functions depend on the acoustic depths of the second ionisation zone of helium and the base of the envelope convective zone, respectively -- through $T_\textrm{He}$ and $T_\textrm{CZ}$ --. 
We observe that the functions depend non-linearly on the values of $T_\textrm{He}$ and $T_\textrm{CZ}$. To preserve the linearity of the method, it is necessary to provide values for $\tau_\textrm{He}$ and $\tau_\textrm{CZ}$ and leave them unchanged to generate the basis and project the observed frequencies over it. In the case of a theoretical model, this estimation is done using the definition of the acoustic depth, i.e.:
\begin{equation}
\tau_\textrm{He/CZ}~=~\int^{r_\textrm{He/CZ}}_{R_*}\frac{dr}{c(r)},
\end{equation}
where $R_*$ is the radius at the photosphere, $r_\textrm{He/CZ}$ represents the radius of the helium second ionisation zone or of the base of the envelope convection zone (in practice we take the corresponding local maximum between the two local minima of $\Gamma_1$ due to the partial ionisation of He and H for $r_\textrm{He}$ and the last point below the surface for which $\nabla < \nabla_\textrm{rad}$ for $r_\textrm{CZ}$).

For observed data, we first generate the optimal model that does not take the glitches into account. We then fit this model for the glitches and retrieve the model values of the acoustic depths as estimators of the optimal values. Therefore, we do not provide any new means to estimate the involved acoustic depths.

Eventually, we may optimise over the values of $\tau_\textrm{He}$ and $\tau_\textrm{CZ}$ to get the best results. This is done through the use of \citet{Brent}'s minimisation algorithm. However, it makes the problem non linear again. Furthermore, the optimised estimations of $\tau_\textrm{He}$ and $\tau_\textrm{CZ}$ always remain very close to the theoretical value and do not decrease significantly the $\chi^2$ value. Also, we observed that by using an initial value of $\tau_\textrm{He}$ different from that at the $\Gamma_1$ maximum (e.g. at the minimum of the helium second ionisation zone) and adjusting it, we found back the value at the maximum. Moreover, we tried to find the value of $Y_f$ giving the best agreement with the helium amplitude (defined in Sect.~\ref{Sec:Ahe}) observed in the case of 16 Cyg A at fixed values of $\tau_\textrm{He}$ corresponding either to the second local minimum or the local maximum of the $\Gamma_1$ profile. We noted that the difference between both values of $Y_f$ is smaller than the standard deviation. In addition, freeing $\tau_\textrm{He}$ does not impact the surface helium abundance retrieved in a significant way. This stems from the fact that the acoustic depth value changes at most of $10~\%$ and has little influence on the measured helium glitch amplitude. Therefore, the influence on the calculated $Y_f$ is negligible as well. Thus we finally decided to give up that last non-linear minimisation for theoretical models. 

By keeping a fully linear implementation of the spectrum fitting we guarantee the stability of the algorithm, the independence\footnote{Provided that the measurements of the frequencies are independent, which is not always the case.} of the parameters obtained via the projection on the orthonormal basis as well as small computation times. This is essential for it has to be included into a non-linear routine that searches for a stellar model accounting at best for the seismic and non-seismic observables.

\section{Seismic indicators}\label{Sec:SeiInd}
The main advantage of the developed method is that it provides -- via Gram-Schmidt's process -- fitted coefficients which are independent of each other. It therefore allows us to derive seismic indicators as uncorrelated as possible. We will define the ones that we explored in the current section. To characterise those indicators, we computed their evolution along the grid of models presented in the next subsection and using the set of modes observed for 16 Cyg A.

\subsection{Models}\label{Sec:Mod}
The grid of models we used was computed using CLES \citep{2008Ap&SS.316...83S} combined with LOSC \citep{2008Ap&SS.316..149S} stellar evolution and oscillation codes. The models used the FreeEOS software \citep{2003ApJ...588..862C} to generate the equation of state table, the reaction rates prescribed by \citet{2011RvMP...83..195A}, the metal mixture of AGSS09 \citep{2009ARA&A..47..481A}, and the OPAL opacity table \citep{1996ApJ...464..943I} combined with that of \citet{2005ApJ...623..585F} at low temperatures. Moreover, the mixing inside convective regions was computed according the mixing length theory \citep{1968pss..book.....C} and using the value $\alpha_\textrm{MLT}~=~l/H_p=~1.82$ (where $l$ is the mixing length and $H_p$ the pressure scale height) that we obtained via a solar calibration. Microscopic diffusion was taken into account in the computation by using \citet{1994ApJ...421..828T}'s routine. For each model, the temperature at the photosphere and the conditions above the photosphere are determined by using an Eddington $T(\tau)$ relationship.
The models have masses ranging from $0.90 M_{\odot}$ to $1.30 M_{\odot}$ by steps of $0.01 M_{\odot}$ and are in the main sequence phase. Moreover, each model has an initial composition of $Y_0 = 0.25$ and $Z_0 = 0.016$ to remain close to the solar case.
Finally, unless specified otherwise, the observed frequencies have been corrected for the surface effects using \citet{2008ApJ...683L.175K}'s prescription of which the coefficients $a$ and $b$ have been calibrated by \citet{2015A&A...583A.112S}.

\subsection{Smooth component indicators}
\subsubsection{Large separation}
A commonly used indicator is the large separation which holds a local (i.e. based on the individual frequencies) and an asymptotic definition. To construct an estimator of the large separation, we will take inspiration in the asymptotic definition. In the asymptotic regime ($n~>>~l$), equation (\ref{Eq:AsyFre}) is satisfied. We notice that, in this formulation, $\Delta$ represents the slope in $n$ of the straight line fitting at best the frequencies.
Moreover, to fit the spectrum smooth component, we project the frequencies over the basis in a specific order given by the sequence of degrees $\left(0,1,2\right)$. This means that keeping only the expression of order $0$ will give an adjustment of the frequencies by a constant term, thus estimating the mean value. Furthermore, if we now keep the expression of first order, we adjust the frequencies to a straight line of which the slope, if we rely on Eq.~\ref{Eq:AsyFre}, is $\Delta$. This is the most common way to define the mean large separation in seismic analyses.
However, we must note that we have different basis vectors depending on the spherical degree considered. This means that, for each value of $l$, we will have a different regression to a straight line, therefore a different estimate of the large separation $\Delta_l$.
Using expressions (\ref{Eq:Bas}), (\ref{Eq:FreFit}), (\ref{Eq:bSmo}), and (\ref{Eq:aSmo}), we isolate this slope to write:
\begin{equation}\label{Eq:Dnul}
\Delta_l~=~a_{l,1} R^{-1}_{l,1,1}.
\end{equation}
We may finally average these indicators over $l$ to estimate at best the large separation. Knowing that the standard deviation of $a_{l,1}$ is $1$, $(R^{-1}_{l,1,1})^2$ is the variance of $\Delta_l$. The weighted mean of the large separations thus yields: 
\begin{equation}\label{Eq:Dnu}
\Delta~=~\frac{\sum\limits_l a_{l,1}/R^{-1}_{l,1,1}}{\sum\limits_l 1/(R^{-1}_{l,1,1})^2}.
\end{equation}

\noindent Finally, we expect from \citet{1986ApJ...306L..37U} that $\Delta$ should be an estimator of the mean stellar density.

\subsubsection{Normalised small separation}
Two commonly used indicators are the small separations $d_{01}(n)$ and $d_{02}(n)$ of which the definitions are:
\begin{eqnarray}
d_{01}(n) & = & \left(\nu(n-1,1)-2\nu(n,0)+\nu(n,1)\right)/2, \label{Eq:d01} \\
d_{02}(n) & = & \left(\nu(n,0)-\nu(n-1,2)\right).\label{Eq:d02}
\end{eqnarray}
They allow a measurement of the spacing between the observations and the asymptotic relation (\ref{Eq:AsyFre}). However, they happen to be sensitive to the surface effects. Therefore, \citet{2003A&A...411..215R} suggested to divide these expressions by the large separation in order to minimise such effects. Indeed, they showed that those ratios are almost independent of the structure of the outer layers of the star.

\begin{figure}[ht]
\centering
\includegraphics[width=0.85\linewidth]{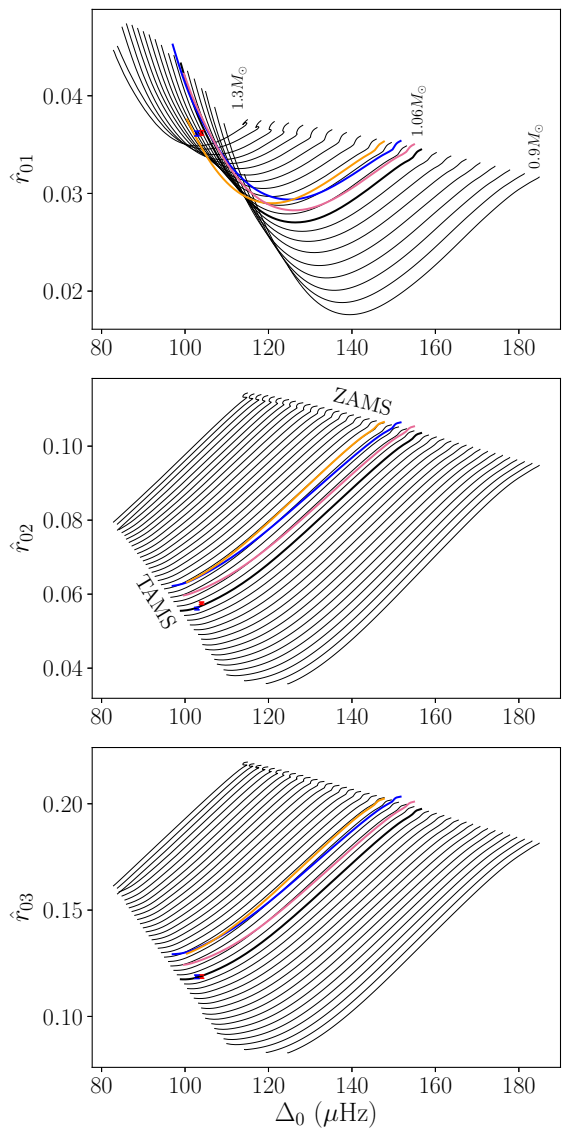}
\caption{Seismic HR diagram defined with the new indicators $\hat{r}_{0l}$ and computed along the grid presented in Sect.~\ref{Sec:Mod}. The masses increase from right to left. The blue marker shows the observed value for 16 Cyg A while the red one shows the value corrected for the surface effects following \citet{2008ApJ...683L.175K}'s prescription. The thick line represents the track for $1.06M_{\odot}$. The blue line has been computed for $Y_0 = 0.27$, the orange one for $\alpha_\textrm{MLT} = 1.5$, and the pink one for $(Z/X)_0 = 0.018$. All the coloured lines have been computed for $1.06M_{\odot}$.}\label{Fig:r0l}
\end{figure}

We thus introduce estimators of these ratios. Such ratios represent the spacing between ridges of spherical degrees $0$ and $1$ for Eq.~\ref{Eq:d01} and degrees $0$ and $2$ for Eq.~\ref{Eq:d02} in the \emph{échelle} diagram \citep{1983SoPh...82...55G}. In a more general way, we approximate the mean difference between the ridges of spherical degrees $0$ and $l$ by comparing the mean values of the frequencies for those degrees. That is $\left(\overline{\nu_0}-\overline{\nu_l}\right)/\Delta_0$. Assuming expression (\ref{Eq:AsyFre}) to be exact, this difference is $\overline{n_0} + \epsilon_0 - (\overline{n_l} + \epsilon_l + l/2)$. Then, we added $-\overline{n_0} + \overline{n_l} + l/2$ to the expression to make its value come close to $\epsilon_0-\epsilon_l$. We then obtained the following expression:

\begin{equation}
\hat{r}_{0l}~=~\frac{\overline{\nu_0}-\overline{\nu_l}}{\Delta_0}+\overline{n_l}-\overline{n_0}+\frac{l}{2},\label{Eq:r0l}
\end{equation}
where $\overline{\nu_l}$ and $\overline{n_l}$ are respectively the weighted mean values of $\nu(n,l)$ and of $n$ for the spherical degree $l$ in accordance with the definition of the scalar product. In addition, the mean value $\overline{\nu_l}$ equals $a_{l,0}R^{-1}_{l,0,0}$ as it is the fitting of the frequencies of degree $l$ to a constant value.
Finally, it has to be stressed that the above expression is slightly different from Eqs. (\ref{Eq:d01}) and (\ref{Eq:d02}) as they represent the local spacing between ridges in the \emph{échelle} diagram and Eq. \ref{Eq:r0l} corresponds to the mean spacing.
Fig.~\ref{Fig:r0l} shows the evolution of those indicators along the grid presented in Sect.~\ref{Sec:Mod}\footnote{However, in the case of the indicator $\hat{r}_{01}$ we used a $0.02M_{\odot}$ step for a better visibility.} for the set of modes observed in 16 Cygni A. The x-axis is the large separation of spherical degree $0$ as we defined above and the y-axis the considered indicator.
We also display the observed values for 16 Cygni A  (\object{HD 186408}) using the frequencies determined by \citet{2015MNRAS.446.2959D}. In blue is the observed value and, in red, the value corrected for the surface effects according to \citet{2008ApJ...683L.175K}'s prescription. We note that these indicators are almost insensitive to surface effects excepted for the case of $\hat{r}_{02}$, which value is changed by about $1\sigma$. We only show the standard deviation for the estimators of the small separation as the one for the large separation is too small to be visible on the plot. Indeed, we computed a standard deviation for $\Delta_0$ of $\sigma\left(\Delta_0\right) = 5~10^{-3} \mu Hz$.

To provide a comparison, we computed the evolution of the `usual' indicators along the same tracks as in Fig.~\ref{Fig:r0l} and display it in Fig.~\ref{Fig:r0lRox} of Appendix~\ref{Ap:Comp}. We observe that the new indicators exhibit the same behaviour as the usual ones and provide smaller standard deviations, therefore, tighter constraints.

We note on Fig.~\ref{Fig:r0l} that $\hat{r}_{02}$ is a very good indicator of the core conditions and should hold information about the evolutionary stage on the main sequence as its evolution is almost monotonic. It is therefore very similar to the small separation that has been shown to carry information on the evolution \citep{1988IAUS..123..295C} as it is sensitive to the sound speed gradient which in turn is sensitive to the chemical composition changes. On the other hand, $\hat{r}_{01}$ is not a good indicator of the evolution but carries additional information. For example, \citet{2010A&A...523A..54D} showed that, for stars with masses and metallicities close to that of $\alpha$ Centauri A (\object{HD128620}), it should provide an upper limit on the amount of convective-core overshooting. We also draw attention to the fact that $\hat{r}_{01}$ shows a turn off for evolved stars. This means that we have to be cautious when fitting models to the observations as, for specific sets of input physics, there exists inaccessible regions. This should allow to constrain the input physics. 

Finally, $\hat{r}_{03}$ does not provide new information.
In addition, we may observe from the comparison between the observations and the theoretical tracks that the expected masses retrieved from the different indicators are in agreement. Indeed, from $\hat{r}_{01}$, we should expect masses between $1.06 M_{\odot}$ and $1.11 M_{\odot}$. Then, from $\hat{r}_{02}$, the expected values are in between $1.06 M_{\odot}$ and $1.07 M_{\odot}$. Again, $\hat{r}_{03}$ does not add some information as the values range from $1.05 M_{\odot}$ to $1.06 M_{\odot}$. As a consequence, we would expect the mass of 16 Cygni A to be around $1.06 M_{\odot}$. This value has been highlighted on the figure by using a thick line. 
However, we must observe that this does not provide a precise estimate of 16 Cyg A mass as we only tested a specific chemical composition -- $Y_0 = 0.25$, $Z_0 = 0.016$ -- as well as given choice of $\alpha_\textrm{MLT} = 1.82$. A proper adjustment is needed to draw conclusions. Those values are only given to illustrate the compatibility between the different indicators. 
We also show the influence of the composition and $\alpha_\textrm{MLT}$ on the coloured tracks. The blue line is for a higher initial helium abundance. Then, the orange line depicts the influence of a lower value of $\alpha_\textrm{MLT}$. Finally, the pink line shows how a higher metallicity modifies the results. We observe in all cases that the inferred mass should be lower.

\subsubsection{$\Delta_{0l}$ indicators}
As it has been shown in several studies, the combination of the small separation ratios $r_{01}$ and $r_{10}$ first introduced by \citet{2003A&A...411..215R} into $r_{010}$ allows to provide inferences about the stellar central mixed region extension \citep{2005AcA....55..177P,2010A&A...514A..31D,2011A&A...529A..63S}. Indeed, the mean value and slope of this indicator occupy very specific regions in the parameter space according to the extent of the central mixed region. This should therefore provide constraint on the amount of overshooting necessary to reproduce observations. As an example, an extensive study of several \emph{Kepler} targets has been realised by \citet{2016A&A...589A..93D} who have been able to provide constraints on the overshooting parameter $\alpha_\textrm{ov}$ for eight of those targets.
In the framework of this paper, the indicator $\hat{r}_{01}$ represents an estimator of the mean value of $r_{010}$. We may therefore build an indicator for its slope as follows:
\begin{equation}
\Delta_{0l}~=~\frac{\Delta_l}{\Delta_0}-1,
\end{equation}
with $\Delta_l$ the large separation for modes of spherical degree l defined by Eq.~(\ref{Eq:Dnul}). It is straightforward to show that, in the asymptotic regime, $\Delta_{01}$ indeed represents the slope of the frequency ratio $r_{01}$.

\begin{figure}
\centering
\includegraphics[width=0.85\linewidth]{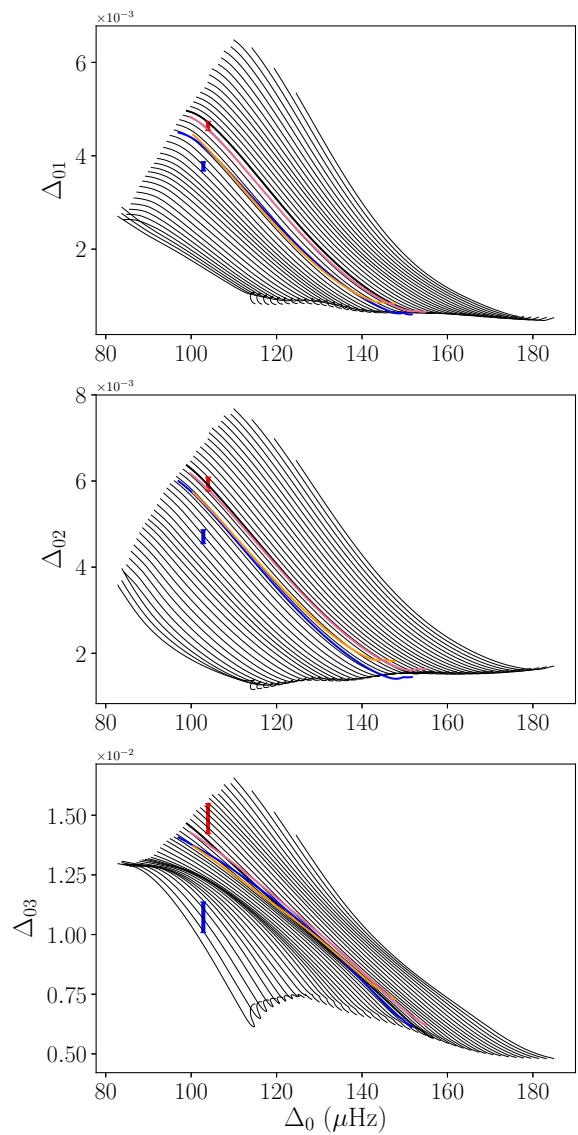}
\caption{Seismic HR diagram defined with the new indicators $\Delta_{0l}$ and computed along the grid presented in Sect.~\ref{Sec:Mod}. The masses increase from right to left. The red marker shows the observed value corrected for the surface effects following \citet{2008ApJ...683L.175K}'s prescription. The thick line represents the track for $1.06 M_{\odot}$. The blue line has been computed for $Y_0 = 0.27$, the orange one for $\alpha_\textrm{MLT} = 1.5$, and the pink one for $(Z/X)_0 = 0.018$. All the coloured lines have been computed for $1.06M_{\odot}$.}\label{Fig:DelRap}
\end{figure}

As for the small separation indicators, we computed the evolution of such indicators along the grid of models presented in Sect.~\ref{Sec:Mod} to demonstrate their regularity and validity. This is shown in Fig. \ref{Fig:DelRap}. We also display the observed value for the case of 16 Cygni A. This value is corrected for the surface effect using \citet{2008ApJ...683L.175K}'s prescription. As in Fig.~\ref{Fig:r0l}, the error on $\Delta_0$ is too small to be visible. Moreover, we show the influence of a change in the composition and of $\alpha_\textrm{MLT}$ with the coloured tracks. The colours are the same as in Fig.~\ref{Fig:r0l}.

Again, it is possible to get an estimate of the mass value for the given composition and physics. From $\Delta_{01}$, we expect masses ranging from $1.06 M_{\odot}$ to $1.07 M_{\odot}$.Then, from $\Delta_{02}$, we expect that they lie between $1.05 M_{\odot}$ and $1.07 M_{\odot}$. Finally, from $\Delta_{03}$, the mass should be between $1.00 M_{\odot}$ and $1.05 M_{\odot}$.
This time, we observe a slight incompatibility between the first two indicators and the last one. Let us add that we highlighted the value of $1.06 M_{\odot}$ as in Fig.~\ref{Fig:r0l}. Moreover, we note that, as opposed to the small separation ratio indicators, the relative behaviours of the coloured tracks are different for the three indicators. This could allow to discriminate various choices in the physics of the models considered as they represent different values of $\alpha_\textrm{MLT}$, $Y_0$, and $Z/X_0$ and also to solve the slight mass discrepancy observed.

\begin{figure}
\centering
\includegraphics[width=0.85\linewidth]{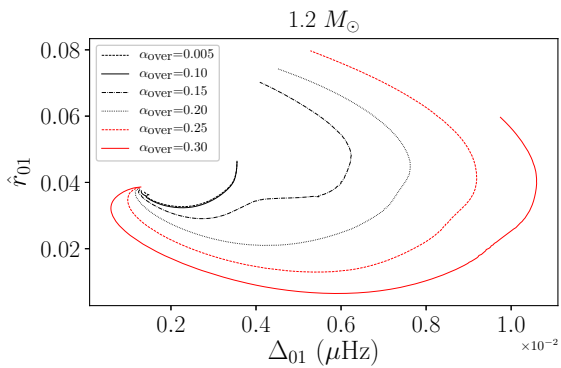}
\caption{MS evolutionary tracks in the ($\Delta_{01}$,$\hat{r}_{01}$) plane for models of $1.2 M_\odot$ with $Y_0 = 0.25$ and $X_0 = 0.734$ for several values of the overshooting parameter $\alpha_\textrm{ov}$ shown in the legend.}\label{Fig:OveDia}
\end{figure}

Finally, as detailed above, the simultaneous use of both $\hat{r}_{01}$ and $\Delta_{01}$ allows to provide estimations of the extent of stellar mixed cores. To illustrate this, we plotted main sequence evolutionary tracks for a $1.2 M_\odot$, $Y_0 = 0.25$ and $X_0 = 0.734$ with several overshooting parameter values ranging from $0.005$ to $0.3$. The overshooting parameter gives the extent of the mixed core above the Schwarzschild limit through $d~=~\alpha_\textrm{ov} min\left(H_p,h\right)$ where $H_p$ is the pressure scale height and $h$ the thickness of the convection zone. This is shown in Fig. \ref{Fig:OveDia}. The ZAMS is at the converging point of the tracks and the TAMS is at their end. We observe that the various tracks occupy very specific regions in the ($\Delta_{01}$,$\hat{r}_{01}$) diagram. Moreover, we note the striking resemblance of Fig. \ref{Fig:OveDia} and Fig. 3 of \citet{2016A&A...589A..93D}. This should allow us to constrain the amount of overshooting.
However, \citet{2016A&A...589A..93D} noted that for this diagnostic tool to be efficient, the mean large separation of the target should not exceed $\sim 110 \mu Hz$ and mixed modes should not be present in the oscillation spectrum.

\subsubsection{$\hat{\epsilon}$}
To provide an estimate of $\epsilon$, we define the following vector subspace, where frequencies are described as:

\begin{equation}
\nu(n,l)~=~\left(n + \frac{l}{2} +\epsilon \right)\hat{\Delta}~=~\left(n + \frac{l}{2}\right)\hat{\Delta} + K,
\end{equation}
where $\hat{\Delta}$ and $K$ are free parameters.

Then, we define an orthonormal basis over this subspace: $\widetilde{\boldsymbol{q}_0}$ and $\widetilde{\boldsymbol{q}_1}$. Finally, by projection of the frequencies over this basis and identification of the several coefficients with the asymptotic formulation, we can retrieve an expression for $\hat{\epsilon}$, the estimator of $\epsilon$. 
We note that this projection also provides an expression to estimate the large separation that is different from Eq.~(\ref{Eq:Dnu}).

\begin{figure}
\centering
\includegraphics[width=0.85\linewidth]{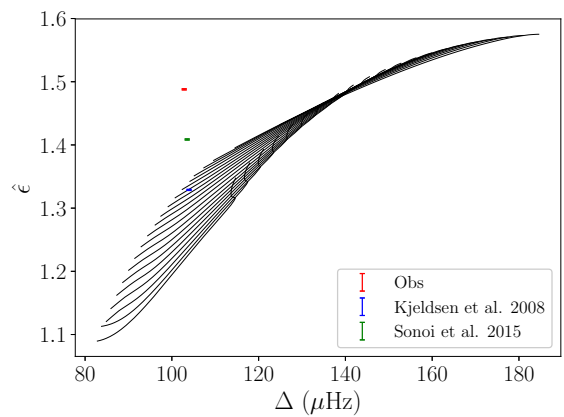}
\caption{Evolution of $\hat{\epsilon}$ along the grid presented in Sect.~\ref{Sec:Mod}. The masses increase from top to bottom. The step is here of $0.02 M_{\odot}$. The red, green and, blue markers respectively represent the observation for 16 Cygni A, the observation for 16 Cygni A corrected for the surface effects using \citet{2008ApJ...683L.175K}'s prescription and the one corrected using \citet{2015A&A...583A.112S}'s prescription.}\label{Fig:EpsEvo}
\end{figure}

Figure~\ref{Fig:EpsEvo} shows the evolution of the indicator $\hat{\epsilon}$ along the grid presented in Sect.~\ref{Sec:Mod} but with a step of $0.02 M_{\odot}$. We may observe that this indicator is almost insensitive to the mass for the early stages of the main sequence. The influence of the mass only becomes visible when the stars become older.
Moreover, the red marker in Fig.~\ref{Fig:EpsEvo} shows the observed value for 16 Cygni A. We note a disagreement between theoretical and observed data. This disagreement can however be tackled by correcting the observed frequencies for the surface effects. This is what the blue and green markers represent. For the blue one, we have computed a correction to the surface effects following \citet{2008ApJ...683L.175K}'s prescription. Then, for the green one, we have computed the correction prescribed by \citet{2015A&A...583A.112S}.  
Therefore, it seems reasonable that the indicator we defined could be of some use to constrain the surface effects. It could provide a complementary method to that of \citet{2016A&A...585A..63R}. Indeed the method presented uses differences between observed and model $\epsilon$ values -- under the hypothesis that both the model and observed star have the same inner structure -- to isolate only the surface contribution to the measured frequencies. This allows to account for surface effects without the need of empirical corrections. On the other hand, the present indicator should allow to discriminate several surface effects corrections without the need of any physical assumption.

\subsection{Glitch indicators}
\subsubsection{Helium amplitude}\label{Sec:Ahe}
With the aim of retrieving the photospheric helium abundance, we built an indicator of the helium glitch amplitude. \citet{2014ApJ...790..138V} obtain their indicator via an integration of the glitch amplitude over the spectrum. We prefer taking advantage of the scalar product and define the indicator as the norm of the helium glitch component, thus generating the following expression:
\begin{equation}
A_\textrm{He}~=~\sqrt{C_\textrm{He,5}^2+S_\textrm{He,5}^2+C_\textrm{He,4}^2+S_\textrm{He,4}^2}.
\end{equation}

\begin{figure}
\centering
\includegraphics[width=0.85\linewidth]{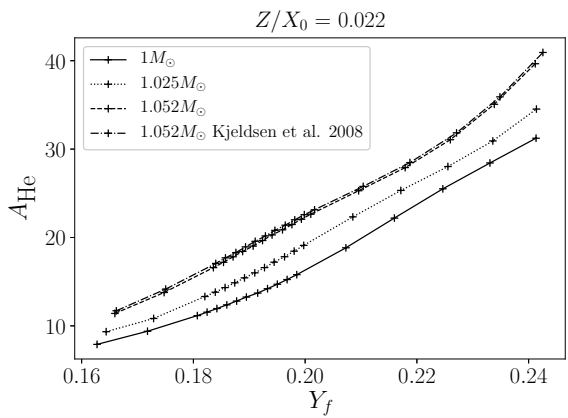}
\caption{Evolution of the helium glitch amplitude $A_\textrm{He}$ with the surface helium abundance $Y_f$. Each track corresponds to a given mass, written in the legend. The dot-dashed line represents the amplitude for a $1.052M_{\odot}$ model of which the frequencies have been corrected for surface effects as in \citet{2008ApJ...683L.175K}. Every model has an initial heavy elements abundance of $(Z/X)_0 = 0.022$. Each point has been computed with the same large separation to remain at the same evolutionary stage.}\label{Fig:AheY}
\end{figure}

\begin{figure}
\centering
\includegraphics[width=0.85\linewidth]{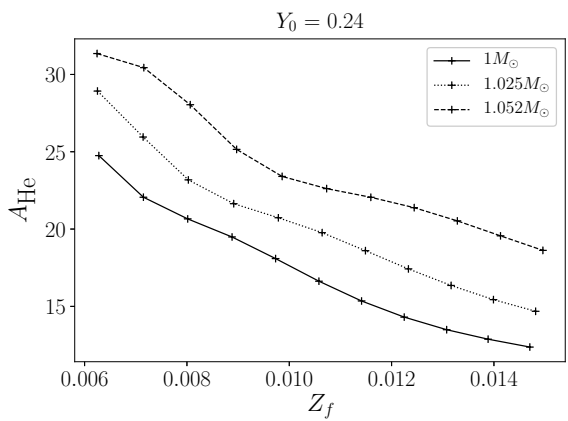}
\caption{Evolution of the helium glitch amplitude $A_\textrm{He}$ with the surface heavy elements abundance $Z_f$. Each track corresponds to a given mass labeled in the legend.}\label{Fig:AheZ}
\end{figure}

Thanks to the orthonormalisation, it is independent of the other indicators and its standard deviation equals $1$.
We calculated the evolution of this indicator with respect to the surface helium mass fraction $Y_f$ and the surface mass fraction of metals $Z_f$. This is shown in Figs.~\ref{Fig:AheY} and \ref{Fig:AheZ}. Fig.~\ref{Fig:AheY} has been computed for stars with a fixed $(Z/X)_0$ ratio of $0.022$ for several surface helium mass fractions, displayed on the abscissa, and for the three values of the mass shown in the legend. To have a reference, we imposed the models to have a fixed value of the large separation -- the one observed for 16 Cyg A.
In a similar way, Fig.~\ref{Fig:AheZ} has been computed for a fixed value of $Y_0 = 0.24$ and for several masses shown in the legend. The helium glitch amplitude has then been computed for the values of $Z_f$ displayed on the abscissa. We insist on the fact that these tracks do not represent the evolution of the helium amplitude with the surface composition along the evolution of a given model. Instead, each point corresponds to a given stellar model that fits the observed 16 Cygni A large separation for a given surface composition. This means that those models were not selected from the grid presented in Sect.~\ref{Sec:Mod}.

We observe in Fig.~\ref{Fig:AheY} an increasing trend in the helium glitch amplitude with the helium mass fraction as well as with the mass. This has to be expected as a larger quantity of helium inside the star would lead to a more important depression of the first adiabatic index $\Gamma_1$ at the second ionisation zone of helium, and, therefore, a glitch of greater amplitude. Moreover, we also show the influence of the surface effects by computing the amplitude evolution for a $1.052 M_{\odot}$ star for which the surface effects have been taken into account via \citet{2008ApJ...683L.175K}'s prescription (dot-dashed line). It is apparent that they have little influence on the amplitude as the values remain in the $1\sigma$ error bars of the uncorrected models. This was expected as the glitch is of greater amplitude in the low frequencies regime while the surface effects corrections are greater in the high frequencies regime.

Furthermore, as shown in Fig.~\ref{Fig:AheZ}, the glitch amplitude and the metallicity are anti-correlated. This corroborates \citet{2004MNRAS.350..277B}'s observations.  Therefore, we are facing a degeneracy and the glitch amplitude alone will not be sufficient to estimate properly the surface helium mass fraction. Thus, the smooth component indicators defined above will be of great help. This clearly shows that the $A_\textrm{He}$-$Y_f$ relation is model dependent which should never be forgotten.

\begin{figure}
\centering
\includegraphics[width=0.85\linewidth]{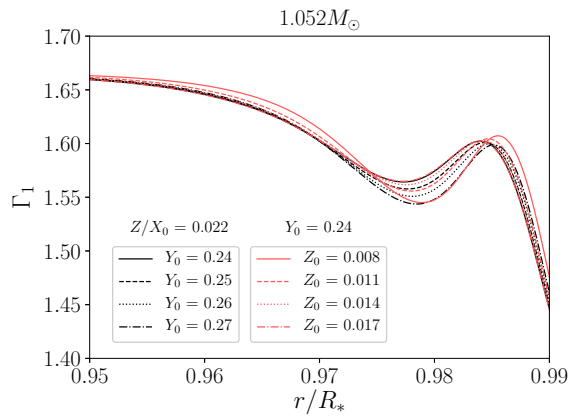}
\caption{Evolution of $\Gamma_1$ as a function of the reduced radius in the superficial layers of stars of fixed large separation. Every star has a mass of $1.052 M_{\odot}$. Two families of curves are displayed. The black ones have a fixed value of the ratio between the initial hydrogen and heavy elements abundances of $(Z/X)_0 = 0.022$ and variable initial helium abundance. The red ones have a fixed initial helium abundance of $Y_0 = 0.24$ and a variable initial heavy elements abundance. The different values are displayed on the figure.}\label{Fig:Gam1r}
\end{figure}

Fig.~\ref{Fig:Gam1r} illustrates this degeneracy. It represents the profile of $\Gamma_1$ as a function of the reduced radius in the superficial layers of stars of a fixed large separation but with several chemical compositions. We immediately notice that both an increase of the surface helium abundance and a decrease of the surface heavy elements abundance lead to a minimum that increases in magnitude. Therefore, the helium glitch amplitude becomes greater as well. We provide an interpretation of this phenomenon in Sect.~\ref{Sec:Lim}.

\subsubsection{Convection zone amplitude}
The definition of the envelope convective zone glitch amplitude we provide is very similar to that of the helium glitch and is the following:
\begin{equation}
A_\textrm{CZ}~=~\sqrt{C_\textrm{CZ}^2+S_\textrm{CZ}^2}.
\end{equation}

\begin{figure}
\centering
\includegraphics[width=0.85\linewidth]{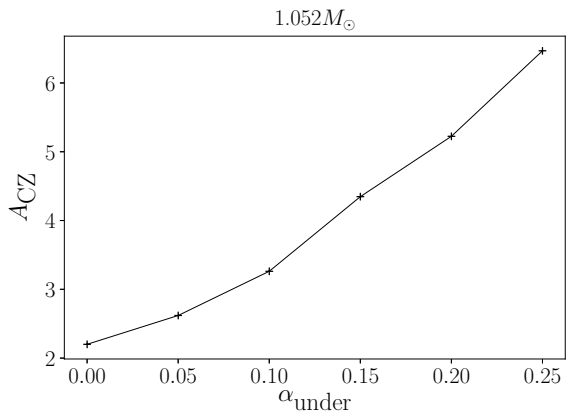}
\caption{Evolution of the convection zone glitch amplitude $A_\textrm{CZ}$ with $\alpha_\textrm{under}$ for a star of mass $1.052 M_{\odot}$, $X_0~=~0.744$, $Z_0~=~0.016$ and at a constant value of the large separation.}\label{Fig:AczAlp}
\end{figure}

We expect this indicator to be a proxy of the sharpness of the transition between the envelope convective zone and the radiative zone. Again, thanks to the orthonormalisation, it is independent of the other indicators and its standard deviation is equal to $1$.
We present in Fig.~\ref{Fig:AczAlp} its evolution with the importance of the undershooting, characterised by the coefficient $\alpha_\textrm{under}$. This coefficient determines the size $d$ of the undershooting region at the bottom of the convective envelope. This size is given by $d~=~\alpha_\textrm{under} min\left(H_p,h\right)$ where $H_p$ is the pressure scale height and $h$ the thickness of the convection zone. In the undershoot region, the temperature gradient is set to the adiabatic one and the mixing is assumed to be instantaneous.

We expect an increase of $A_\textrm{CZ}$ with $\alpha_\textrm{under}$ as the introduction of undershooting in a stellar model will create a discontinuity of the temperature gradient. The temperature gradient jump increases with the value of $\alpha_\textrm{under}$. Thus, the glitch amplitude increases as well.
This is what we show in Fig.~\ref{Fig:AczAlp} where the computed models have a mass of $1.052M_{\odot}$, an initial hydrogen mass fraction of $X_0~=~0.744$, and an initial metal mass fraction of $Z_0~=~0.016$. As for the case of the helium amplitude evolution, we kept a fixed large separation -- which is that of 16 Cyg A -- for each model. Moreover, we noted that each computed model was at the same evolutionary stage (constant central hydrogen mass fraction). Therefore, the observed effect is not evolutionary but rather the effect of the temperature gradient discontinuity as expected. Finally, we may add that, when setting the temperature gradient to the radiative one in the undershoot region, we do not observe any significant trend in the amplitude with $\alpha_\textrm{under}$. We may thus conclude that we observe the effect of the temperature gradient and not of the chemical composition.

\section{Method characterisation}\label{Sec:Cha}
\subsection{Capabilities}
\begin{figure}
\centering
\includegraphics[width=0.85\linewidth]{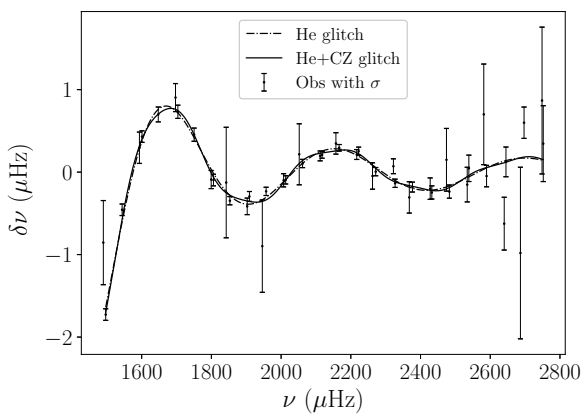}
\caption{Fitted glitch to 16CygA (\object{HD 186408}) data \citep{2015MNRAS.446.2959D}. Only the $l=0$ fitted curves are displayed.}\label{Fig:GliObs}
\end{figure}

For the observed data, we have chosen the frequencies computed by \citet{2015MNRAS.446.2959D} for the component A of the binary system 16 Cygni (\object{HD 186408}). We come back to the particular case of 16 Cyg A in Sect.~\ref{Sec:16Cyg}. Figure~\ref{Fig:GliObs} shows the difference between the observed frequencies and the smooth component of the fitted frequencies, the fitted helium glitch alone and the fitted helium and convection zone glitches. Only the $l=0$ fitted curves are displayed. We observe that the fit is good and that the helium glitch has been properly isolated. However, the convection zone glitch is of very low amplitude, compared to the helium glitch, and has a negligible contribution. This is visible in the negligible improvement of the $\chi^2$ value from the results without including the convective zone glitch (about $10\%$ variation). These results are similar to those of \citet{2014ApJ...790..138V}.

\subsection{Limitations}\label{Sec:Lim}
The presented method has been developed for the study of solar-like pulsators and to provide a comprehensive analysis of their oscillation spectra. However, it is not yet adapted to study evolved stars which exhibit mixed modes.

\begin{figure}
\centering
\includegraphics[width=0.85\linewidth]{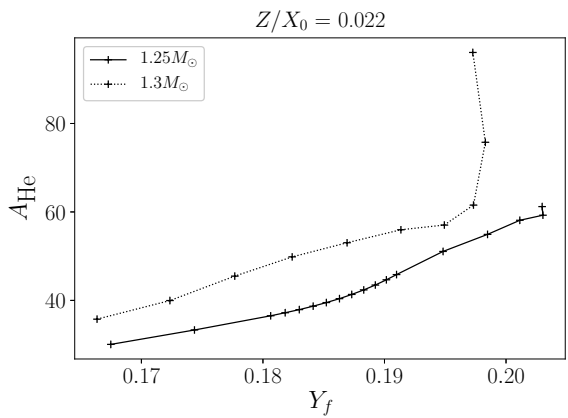}
\caption{Evolution of the helium glitch amplitude $A_\textrm{He}$ with the surface helium abundance $Y_f$ for high masses.}\label{Fig:AheLim}
\end{figure}

In addition, from masses around $1.25M_{\odot}$ and above as well as for the highest values of the helium abundances considered (from $Y_f~\sim~0.195$ and above), the evolution of the helium glitch amplitude with the surface helium mass fraction is less monotonic and inferences become unreliable. Fig.~\ref{Fig:AheLim} shows this limitation. Indeed, for $1.25 M_{\odot}$ we observe a very sharp increase of the helium amplitude as we defined in Sect.~\ref{Sec:Ahe} for a quasi constant surface helium mass fraction. This corresponds to a decrease of the surface $Z/X$ ratio which is shown in Sect.~\ref{Sec:Ahe} to lead to a higher amplitude. 
Also, we observe for $1.3M_{\odot}$ that the last point ($Y_f \sim 0.197$ and $A_\textrm{He} \sim 96$) moves backwards. This is due to microscopic diffusion. We observe that the size of the convective envelope decreases with the initial helium abundance. This leads to a more efficient gravitational settling as the diffusion velocities are greater close to the surface. Therefore, the surface abundance becomes smaller than for the previous point that had a lower initial helium abundance. We observe the same decrease in the surface metallicity which explains the increase of the amplitude even though the surface helium mass fraction remained constant. 

\begin{figure}
\centering
\includegraphics[width=0.85\linewidth]{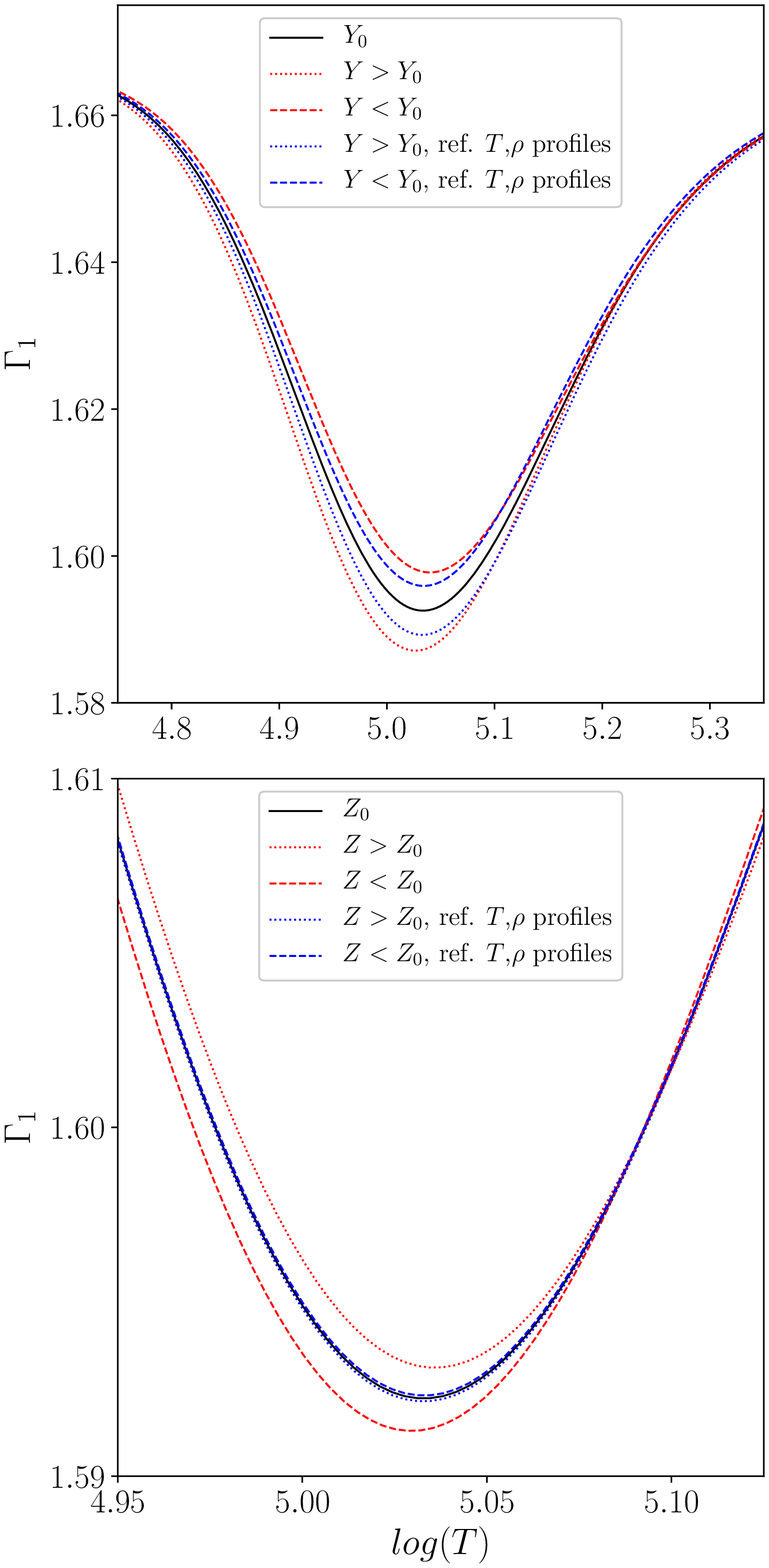}
\caption{Comparison with a reference model (black curve) of the toy model $\Gamma_1$ profiles in cases for which the temperature and density profiles have been decoupled from the composition profiles (blue curves) and coupled cases (red curves). The dashed lines have a common decreased abundance of the considered element and the dotted lines have an increased abundance. The top panel shows the effect of the helium abundance and the bottom panel of the metals abundance.}\label{Fig:GamTM}
\end{figure}

We also noted in Sect.~\ref{Sec:Ahe} that the surface abundance of metals has a relevant influence on the helium glitch amplitude and separating its contribution from that of the helium is not an easy task.
To investigate such a behaviour, we developed a toy model for the first adiabatic index $\Gamma_1$ which is thought to be the main contributor to the helium glitch amplitude \citep{1990LNP...367..283G,2007MNRAS.375..861H}. In this model, we trace back the influence of the helium and metals abundances on the dip of $\Gamma_1$ in the helium second ionisation zone. More information about the construction of this model is given in Appendix~\ref{Ap:Toy}. 
Using it, we were able to test the influence of the chemical composition decoupled from its evolutionary effect on the temperature and density profiles. To do so, we artificially modified the chemical composition profile of a reference model (black curve in Fig.~\ref{Fig:GamTM}), without changing its temperature and density profiles, to match the surface abundance, in either metals or helium, of a second reference model. This way, we were able to isolate the contribution of the chemical composition alone.
Fig.~\ref{Fig:GamTM} shows the comparison between the helium second ionisation zone toy models with modified chemical composition profiles only (blue curves) and toy models with the same composition but for which its effect on temperature and density have been taken into account (red curves). We observe in the top panel that the effect of the helium abundance dominates over the effect of temperature.
In the bottom panel, we observe that the metal abundance alone does not modify by a significant amount the dip in the helium second ionisation zone. However, when taking its influence on the temperature and density profiles into account, the effect becomes significant. This allows to better understand the degeneracy in composition on the helium amplitude. One effect, that of the helium abundance is direct on the shape of the $\Gamma_1$ profile in the helium second ionisition zone while the second, that of the metal abundances, is indirect as it influences the temperature and density profiles which in turn modify the $\Gamma_1$ profile.

Finally, we observe that, in a $lnP~-~lnT$ diagram, the curves for the different models in the temperature region of the $\Gamma_1$ dip are parallel to each other but at various height. A higher $lnP~-~lnT$ curve corresponds to a shallower $\Gamma_1$ dip. We also observe this behaviour with a fixed composition and a variable mass. Higher masses models have lower curves and deeper depressions in $\Gamma_1$. Therefore, the toy model allows to understand the influence of the mass on the helium glitch amplitude as well.

\section{Illustration with 16 Cygni A (\object{HD 186408}) observations}\label{Sec:16Cyg}
16 Cyg A (\object{HD 186408}) is one of the brightest stars in the Kepler field of view. It belongs to a binary system of solar analogs, both exhibiting solar-like pulsations. The quality and length of the collected time series makes it the ideal subject to test the method. It should be noted that we do not provide a detailed study of 16 Cygni A. Indeed, we only present here the capability of the method to provide structural constraints. A detailed study will be the object of a future paper.

\subsection{Methodology}
To obtain constraints on 16 Cyg A, we compute the value of the seismic indicators using the frequencies determined by \citet{2015MNRAS.446.2959D}. We have corrected the surface effects for the observed frequencies by using the power law prescribed by \citet{2008ApJ...683L.175K} and the $a$ and $b$ coefficients fitted by \citet{2015A&A...583A.112S} as a function of $T_\textrm{eff}$ and $g$. The authors have done this coefficient adjustment by comparing the adiabatic frequencies of patched models based on 3D simulations and that of unpatched standard 1D models.
We then fit the observed values of $\Delta$, $\hat{r}_{01}$, $\hat{r}_{02}$, and $A_\textrm{He}$ with the age, mass, initial mass fraction of hydrogen, and the initial heavy elements over hydrogen abundance ratio as free parameters.
To do so, we select an initial guess value of $X_0$ and derive the best fit values of $\Delta$, $\hat{r}_{01}$, and $\hat{r}_{02}$. This results in a set of values for the mass, the age, and $(Z/X)_0$. Then, the adjustment is done by applying the secant method to find the value of $X_0$ in best agreement with the target $A_\textrm{He}$. At each step of the secant algorithm -- at each value of $X_0$ --, \emph{Levenberg-Marqardt}'s algorithm (L-M) computes the optimal set of stellar parameters giving the best fit of $\Delta$, $\hat{r}_{01}$, and $\hat{r}_{02}$.
Finally, when we have a good estimate for $A_\textrm{He}$, we use L-M's algorithm one last time to derive the complete set of values for the mass, the age, $X_0$, and $(Z/X)_0$ fitting the observed parameters. We used the secant method in order to diminish the computational time needed to converge towards the solution.

\subsection{Results}\label{Sec:Res}
\begin{figure}
\centering
\includegraphics[width=0.85\linewidth]{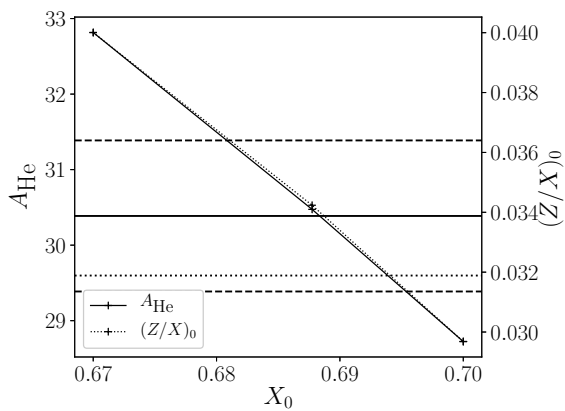}
\caption{Helium amplitude (solid line) and $(Z/X)_0$ ratio (dotted line) versus the initial hydrogen abundance. The horizontal solid line represents the target value for the amplitude -- corrected for the surface effects following \citet{2008ApJ...683L.175K} -- while the dashed lines represent the $1\sigma$ interval. The horizontal dotted line is the value of the amplitude without surface effects correction.}\label{Fig:AHeX}
\end{figure}

\begin{figure}
\centering
\includegraphics[width=0.85\linewidth]{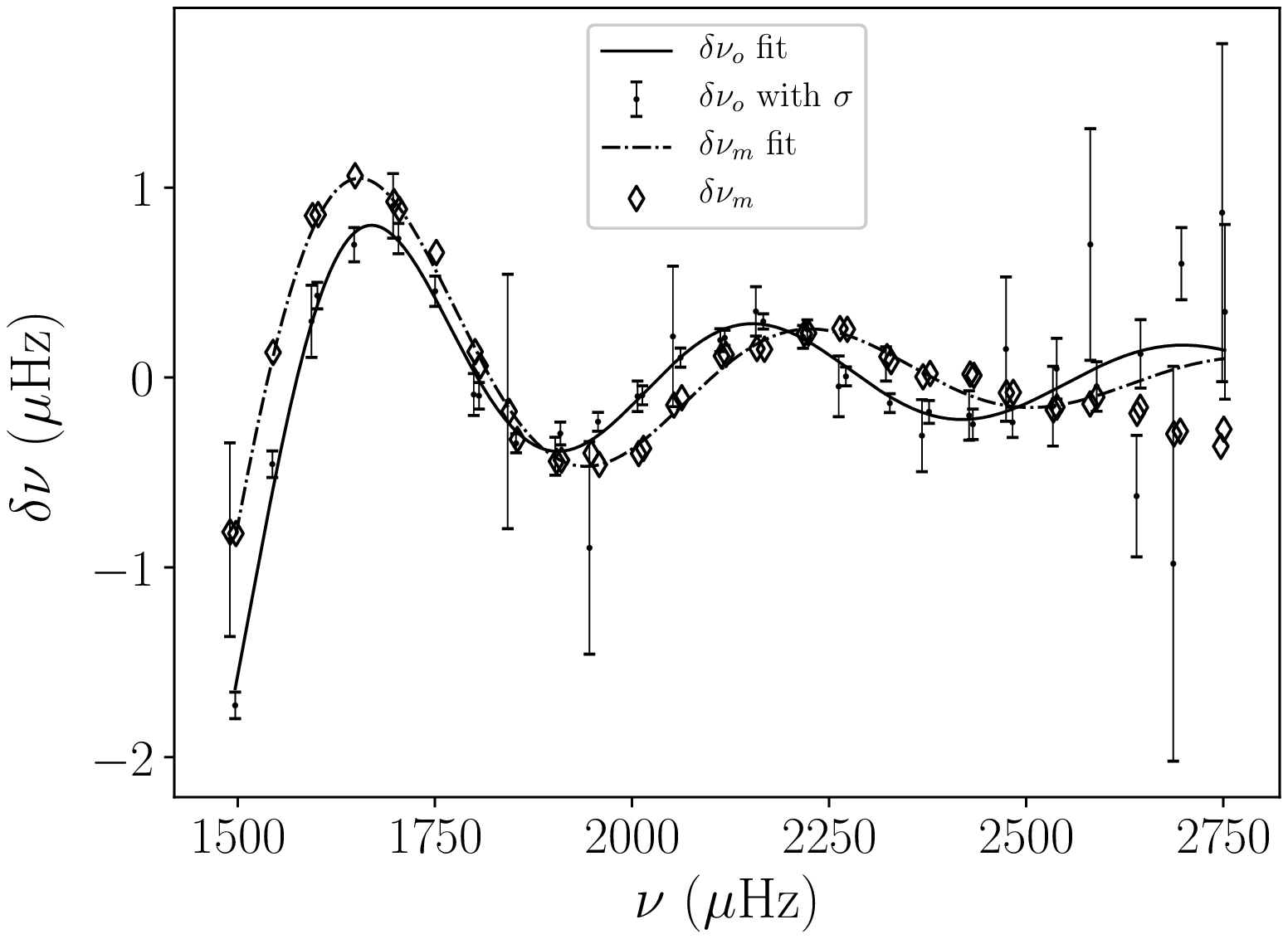}
\caption{Comparison between the observed helium glitch $\delta\nu_o$ (solid line) and the one resulting from the best fit model $\delta\nu_m$ (dot-dashed line), for $l=0$. We also display the observed glitch as a function of the frequencies (errorbars) as well as the best model glitch associated with the theoretical frequencies (diamond).}\label{Fig:ObsFit}
\end{figure}

\begin{figure}
\centering
\includegraphics[width=0.85\linewidth]{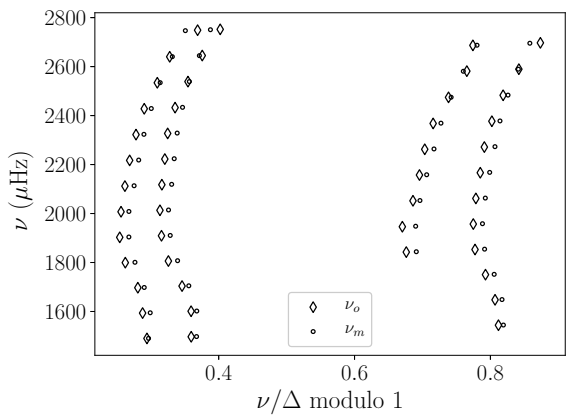}
\caption{Comparison between the observed frequencies (diamonds) and the best model frequencies (circles) in an \emph{échelle} diagram.}\label{Fig:Ech}
\end{figure}

The values of the relevant seismic indicators as well as the associated standard deviations computed from the observed frequencies are given in Table~\ref{Tab:SeiInd}. The column labeled \emph{Kjeldsen} refers to the observed indicators corrected for the surface effects using \citet{2015A&A...583A.112S}'s coefficient fitted to \citet{2008ApJ...683L.175K}'s prescription as explained in Sect. \ref{Sec:Mod}. The \emph{Sonoi} column corresponds to the correction of surface effects using a Lorentzian profile as in \citet{2015A&A...583A.112S}. We also derived the seismic indicators while adding smooth basis elements with $n^{-1}$. As the frequencies are projected in a specific order to build the seismic indicators, only the glitches amplitudes are affected and we have: $A_\textrm{He}~=~27.6$ and $A_\textrm{CZ}~=~3.3$.
Using the methodology described above, we managed to derive the stellar parameters given in Table~\ref{Tab:StePar} for the different sets of observed seismic indicators. We note that changing the treatment of surface effects or adding the $n^{-1}$ basis elements have an impact on the fitted parameters comparable to that of the frequencies uncertainties. Thus, we will only discuss here the parameters adjusted to the data corrected for the surface effects using \citet{2008ApJ...683L.175K}'s prescription and without adding $n^{-1}$ basis elements.
The parameters we derived do not constitute a detailed characterisation of the target 16 Cyg A. Rather, they illustrate the ability of the method to provide constraints on a solar analog. Indeed, only one single set of input physics was tested. Therefore, the standard deviations tend to be underestimated as they are the ones intrinsic to the method. The abundances used for the computations were the solar ones determined by \citet{2009ARA&A..47..481A}.
Let us add that we obtain a surface helium abundance of $Y_f~=~0.242~\pm~0.028$ which lies in the interval obtained by \citet{2014ApJ...790..138V}, $Y_{f,V} \in \left[0.231,0.251\right]$. This is comforting us in the idea that the developed method is efficient in isolating the glitches and drawing inferences from their signatures.
In Fig.~\ref{Fig:AHeX}, we show the evolution of the helium glitch amplitude resulting from the 3 parameters adjustment of $\Delta$, $\hat{r}_{01}$, and $\hat{r}_{02}$ as a function of $X_0$. We also show the evolution of the value of $(Z/X)_0$. The very linear trend justifies that we used the secant method to provide successive estimates of $A_\textrm{He}$ in order to lessen the computational charge. In addition, we illustrate both the observed value for $A_\textrm{He}$ and the corrected value under \citet{2008ApJ...683L.175K}'s prescription. We note that the corrected value is of about $0.79\sigma$ lower than the uncorrected one. Using \citet{2015A&A...583A.112S}'s prescription only leads to a $0.55 \sigma$ variation of the measured amplitude. This demonstrates that the surface effects have a small influence on the amplitude we derive using the method.
Also, we retrieved a value of $970.97~s$ for $\tau_\textrm{He}$ and $3042.32~s$ for $\tau_\textrm{CZ}$. As they were fixed to model values and, as a consequence, were neither parameters nor constraints of the adjustment, we do not provide uncertainties. However, we may compare their values to the ones adjusted by \citet{2014ApJ...790..138V}. They obtained $\tau_\textrm{He} \in \left[ 868, 944 \right]~s$ and $\tau_\textrm{CZ} \in \left[ 2992, 3234 \right]~s$. We observe that $\tau_\textrm{CZ}$ lies in the interval calculated by \citet{2014ApJ...790..138V} while the value of $\tau_\textrm{He}$ is slightly above the upper limit. However, as we show in Sect.~\ref{Sec:MetGli} this does not impact the inferences drawn from the helium glitch amplitude in a significant way. To illustrate this statement, we freed the value of $\tau_\textrm{He}$ (but not that of $\tau_\textrm{CZ}$ as the convection zone glitch is of negligible amplitude compared to that of the helium glitch). The relative change between the optimised $\tau_\textrm{He}$ and its estimator is of only $6~\%$. Moreover, the observed value of the helium glitch amplitude remains unchanged compared to its standard deviation (we observe a change of $0.06~\sigma$). As expected, the best fit $Y_f$ also remains untouched. At this point, one should be reminded that we focus on the glitch amplitude to draw our inferences. And, as we showed in Sect.~\ref{Sec:MetGli}, the exact location of the glitch does not have a significant impact on its amplitude as well as on the derived surface helium abundance.
Finally, Fig.~\ref{Fig:ObsFit} shows the fitted helium glitches\footnote{We do not display the envelope convective zone glitch as its amplitude is negligible compared to that of the helium glitch, as mentioned in Sect.~\ref{Sec:Cha}.} for the observed data and the 4 parameters best fit model. It is visible that the observed and fitted glitches are close in amplitude and period. This demonstrates the ability of the method to both isolate the glitch and the related parameters and to provide a model reproducing at best those parameters. We also show in Fig.~\ref{Fig:Ech} the comparison in an \emph{échelle} diagram of the observed frequencies and those of the best fit model. We observe in this last figure that the smooth component of the spectrum (not visible in the previous figure) is properly adjusted. Thus, by fitting the set of indicators $\Delta_0$, $\hat{r}_{01}$, $\hat{r}_{02}$, and $A_\textrm{He}$ we obtain a good representation of the observed frequencies. We also show in App. \ref{Ap:SupAdj} the adjustment to frequencies corrected for surface effects using a Lorentzian profile as in \citet{2015A&A...583A.112S}. We observe in Fig. \ref{Fig:GliSon} that the agreement between the model and observed glitches is better than in Fig. \ref{Fig:ObsFit}. However, we also note in the corresponding \emph{échelle} diagram (Fig. \ref{Fig:EchSon}) that there is an offset between observed and model ridges. This corresponds to different values of $\hat{\epsilon}$. Fig.~\ref{Fig:EpsEvo} also illustrates this behaviour as it is clearly visible that only \citet{2008ApJ...683L.175K}'s prescription allow to reproduce the observed value for $\hat{\epsilon}$.

\begin{table}
\centering
\begin{tabular}{cccc}
\hline
Indicator & \multicolumn{2}{c}{Value} & $\sigma$ \\ 
 & Kjeldsen & Sonoi & \\
\hline
\hline\\[-0.8em]
$\Delta (\mu Hz)$ & $104.088$ & $103.611$ & $0.005$ \\
$A_\textrm{He}$ & $30.4$  & $30.1$ & $1.0$ \\
$A_\textrm{CZ}$ & $2.2$  & $1.5$ & $1.0$ \\
$\overline{\epsilon}$ & $1.3288$ & $1.4086$ & $0.0009$ \\
$\hat{r}_{01}$ & $0.0362$ & $0.0362$ & $0.0002$ \\
$\hat{r}_{02}$ & $0.0575$ & $0.0561$ & $0.0003$ \\
$\hat{r}_{03}$ & $0.1187$ & $0.1184$ & $0.0008$ \\
$\Delta_{01}$ & $4.6~10^{-3}$ & $3.8~10^{-3}$ & $0.1~10^{-3}$ \\
$\Delta_{02}$ & $5.9~10^{-3}$ & $4.8~10^{-3}$ & $0.1~10^{-3}$ \\
$\Delta_{03}$ & $14.9~10^{-3}$ & $10.6~10^{-3}$ & $0.6~10^{-3}$ \\[0.2em]
\hline 
\end{tabular}
\caption{Observed seismic indicators.}\label{Tab:SeiInd}
\end{table}

\begin{table}
\centering
\begin{tabular}{ccccc}
\hline
Quantity & \multicolumn{3}{c}{Value} & $\sigma$ \\ 
 & \multicolumn{2}{c}{Kjeldsen} & Sonoi & \\
 & $n^0,n,n^2$ & $n^{-1}$ &  & \\
\hline
\hline\\[-0.8em]
$M (M_{\odot})$ & $1.06$ & $1.06$ & $1.06$ & $0.02$ \\
$R (R_{\odot})$ & $1.218$ & $1.219$ & $1.223$ & $0.001$ \\
age (Gyr) & $6.8$ & $6.9$ & $7.1$ & $0.1$ \\
$X_0$ & $0.684$ & $0.697$ & $0.685$ & $0.010$ \\
$(Z/X)_0$ & $0.035$ & $0.031$ & $0.036$ & $0.002$ \\
$Y_f$ & $0.242$ & $0.232$ & $0.240$ & $0.028$ \\
$\left[ Fe/H \right]$ & $0.188$ & $0.131$ & $0.199$ & $0.03$ \\[0.2em]
\hline
\end{tabular}
\caption{Adjusted stellar parameters.}\label{Tab:StePar}
\end{table}    

\section{Discussion and conclusions}
\subsection{Principle}
In the present paper, we provide a new method that uses as much of the available seismic information as possible. Indeed, we take advantage of the information contained in both the glitches and the smooth component (usually discarded in glitches analyses but then used separately for forward seismic modelling, see \citet{2014ApJ...790..138V} for example) of the spectrum to define indicators that are as independent as possible of each other. To do so, we take advantage of Gram-Schmidt's algorithm to create an orthonormal basis over which we project the frequencies. The obtained coefficients are therefore independent of each other and consist in a linear combination of the frequencies. Thus, using the appropriate combination of those allows us to define uncorrelated indicators. Such indicators are constructed in order to reproduce the behaviour of `usual' indicators such as the large frequency separation. Up to this day and to our knowledge, no method, has been proposed to provide proper correlations between the smooth and glitch indicators as they are built separately. Therefore, our method provides the asteroseismologists with new diagnosis means.

\subsection{Advantages}
As the method only relies on linear algebra, it is very stable and the computation times are negligible -- of the order of a fraction of a second --. Thus, it could easily be implemented in stellar model fitting algorithms, which are non-linear, without impacting the total computational time. This can be done with any algorithm as the method only focuses on the definition of new seismic indicators and does not rely on the physics of the model itself. Let us add that the defined indicators should be used in combination with non-seismic constraints through a single merit function while searching for a stellar model in order to obtain proper covariances between the inferred quantities -- which is not often the case in seismic analyses. The illustration for 16 Cyg A truly demonstrates the possibility to use the new indicators to provide further constraints on stellar structure in the framework of forward seismic modelling.

In addition, the usual indicators often hold a local definition and may use correlated information while the new ones are built in a way that the information used is averaged over the whole frequency range and that it is not redundant (each observed frequency is used only once for each indicator). This results in smaller standard deviations and smoother behaviours. The Appendix~\ref{Ap:Comp} shows the evolution of the classical indicators $r_{01}(n)$ and $r_{02}(n)$ as defined by \citet{2003A&A...411..215R}. We indeed observe that the new definitions of the indicators give smaller error bars while preserving the expected trends, providing tighter constraints.
Finally, the method has the advantage that it can be implemented even in cases where some modes are missing. It is not the case of the `usual' methods as they need successive modes to define some of their indicators. For example, the classical local definitions of the large separation and second differences require at least two consecutive frequencies. Therefore, whenever some modes are missing, pieces of information might be discarded. 

\subsection{Information carried by the indicators}
We have defined indicators that provide estimates for the classical indicators that are the large separation, $\Delta$ and $\Delta_l$, and the small separation ratios, $\hat{r}_{0l}$. As expected, $\Delta$ provides an estimation of the stellar mean density \citep{1986ApJ...306L..37U}.
Also, we defined the indicators $\Delta_{0l}$ which combine the large separations associated with one specific spherical degree $l$ and which are known to give an estimate of the resonant cavity of the $l$ degree modes \citep{2002ASPC..274...77M}.
Those indicators provide an estimator of the slope of $r_{010}$ which, combined with $\hat{r}_{01}$ can be used to constrain the overshooting parameter as shown in Fig. \ref{Fig:OveDia}.

Moreover, we observe in Fig.~\ref{Fig:r0l} that the indicator $\hat{r}_{02}$, defined to estimate the small separation ratio between the spherical degrees $0$ and $2$, is a proper indicator of the evolution of the star, as expected from \citet{1988IAUS..123..295C}. Also, we observe that the indicator $\hat{r}_{01}$ presents a degeneracy, as a turn-off occurs in the $\Delta_0$-$\hat{r}_{01}$ plane, and there is an inaccessible region. This is a very interesting observation as it should provide tight constraints on the stellar structure. Indeed, we show that a change in composition or in the value of $\alpha_\textrm{MLT}$ allows to modify that region. It will therefore be necessary to use such parameters to reproduce observed values.

Furthermore, Fig.~\ref{Fig:EpsEvo} shows that we might get constraints on the surface effects from the indicator $\hat{\epsilon}$ we defined. It should allow to discriminate from several empirical formulations meant to account for the surface effects. We have tested both formulations from \citet{2008ApJ...683L.175K} and \citet{2015A&A...583A.112S} as an illustration of the diagnosis power of the indicator $\hat{\epsilon}$. However, we are aware of the existence of the formulation from \citet{2014A&A...568A.123B} and it should also be inspected in further studies. Besides, we showed that the helium amplitude indicator is almost unaffected by the surface effects as it has to be expected (see for example Fig.~\ref{Fig:AHeX}). Indeed, empirical surface effects corrections are important for the high frequencies compared to $\nu_\textrm{max}$ while the glitch is of great amplitude only for the low frequencies.
Also, the helium glitch amplitude should allow to draw inferences on the surface helium abundance as shown by Fig.~\ref{Fig:AheY}. However, attention has to be paid as it is also anti-correlated with the metallicity (see Fig.~\ref{Fig:AheZ}). We have demonstrated via a toy model for the first adiabatic index that both effects on the amplitude stem from the position of the adiabat which in turn determines the amplitude of the second local $\Gamma_1$ minimum due to helium partial ionisation.

In addition, we observe that the convective zone glitch amplitude has a significantly lower amplitude than that of the helium glitch and is correlated with the amount of undershooting at the base of the envelope convection zone (in agreement with \citealt{2014ApJ...790..138V}).

Finally, let us add that it is possible to define other indicators than those presented in this paper. This should therefore be carefully studied to take advantage of as much of the available information as possible.

\subsection{Limitations}
However, we show that the method is only fit to draw inferences about solar-like stars, that is low-mass stars on the main-sequence. Indeed, Fig.~\ref{Fig:AheLim} illustrates that from masses around $1.3 M_{\odot}$ and above, diffusion plays an important role and the relation between helium surface abundance and the helium glitch amplitude is not monotonic anymore. Therefore, the method will have to be adapted for massive and evolved stars.

Moreover, one could argue that using model values as estimators of the acoustic depths of the glitches is a major drawback of the method. However, the proposed method does not focus on those quantities. Rather, we focus on the information that the amplitude of the glitch (not the period of the signal) carries. This means that we only need proper, but not exact, estimators for the acoustic depths in order to draw inferences. Indeed, we showed in Sect.~\ref{Sec:MetGli}, in the case of the helium glitch, that a small excursion from the estimated value (either by manually setting another value, that of the second minimum in the first adiabatic index for the helium glitch, or by optimising over its value) does not lead to a significant change in the measured amplitude. Therefore, the inferences drawn remained unchanged. However, one could still regard this as a flaw of the method as we do not provide a new way of retrieving the acoustic depths of glitches. This could be explored in future studies by, for example, finding the global opitmum for the acoustic depths of the glitches and then using the measured of the value as a constraint for the best fit model. Nevertheless, this would make the calculations more time consuming and annihilate the benefit of the orthonormalisation that is the independence of the fitted parameters.

Also, the indicators should be used to complement non-seismic data as we have shown that, for example, there exists a degeneracy between the helium glitch amplitude, the helium surface abundance and the metallicity\footnote{Which we have shown in Sect.~\ref{Sec:Lim} to be an indirect effect as it is due to the influence on density and temperature profiles for the metallicity.} (see Figs.~\ref{Fig:AheY} and \ref{Fig:AheZ}). Therefore, the helium amplitude alone is not sufficient and additional information are needed to lift such a degeneracy -- these information may be contained in other indicators (seismic and non-seismic).

\subsection{Future perspective}
The next step will be the detailed study with the new method of several stars from the \emph{Kepler legacy sample} \citep{2017ApJ...835..172L}. This sample consists of 66 main sequence stars for which at least one year of continuous observations has been made. Having such long time series provides the necessary precision to study glitches. Let us add that this future study is intended to be independent of what has been done in \citet{2017ApJ...837...47V}.
 
Also, another important step will be the improvement of the method by enabling the study of glitches present in red subgiants spectra. The peculiarity of such pulsators is that they exhibit mixed modes. This will consist in a new challenge as their analytical formulation is not a simple task. However, this is a necessary step to improve our knowledge of the evolution of a star such as the Sun. Moreover, this will allow a better description of such stars and will provide a deeper understanding of their properties via, for example, the characterisation of the mixing processes.

\section{Acknowledgments}
The authors would like to thank the referee for their very constructive remarks that improved the paper in a significant way.\\
M.F. is supported by the FRIA (Fond pour la Recherche en Industrie et Agriculture) - FNRS PhD grant.\\
S.J.A.J.S. is funded by ARC grant for Concerted Research Actions, financed by the Wallonia-Brussels Federation. \\
GB acknowledges support from the ERC Consolidator Grant funding scheme ({\em project ASTEROCHRONOMETRY}, G.A. n. 772293).

\bibliographystyle{aa}
\nocite{*}
\bibliography{bibli}

\begin{appendix}
\section{Effect of the regularisation constant}\label{Ap:Lam}
\begin{figure}
\centering
\includegraphics[width=0.85\linewidth]{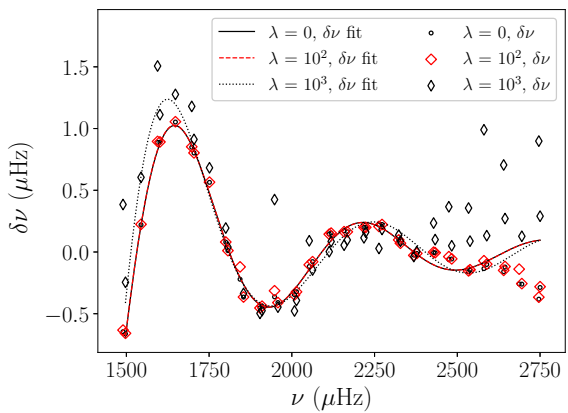}
\caption{Comparison between the glitch adjustment for several values of the regularisation constant $\lambda$ for 16 Cyg A best model determined in Sect.~\ref{Sec:Res}. The values are given in the legend. The lines correspond to the fitted glitch and the markers to the observed glitch. Only the fitted curves for $l=0$ are displayed.}\label{Fig:Lam}
\end{figure}

Fig.~\ref{Fig:Lam} shows the influence of the regularisation constant $\lambda$ on the quality of the fitted glitch. We fitted the glitch of the best model obtained in Sect.~\ref{Sec:Res} using the values of $\lambda$ shown in the legend. We observe that, when using a value of $10^2$ which is already significant as discussed in Sect.~\ref{Sec:Smo}, the fit (red dashed curve) remains close to the one obtained without including regularisation terms (solid black curve). There is only a slightly higher dispersion in the model frequencies subtracted from the smooth part of the spectrum.
To show the degradation caused by a regularisation constant that dominates the adjustment, we used $\lambda = 10^3$. We observe both a discrepancy between results without and with (dotted black curve) regularisation terms and a higher dispersion. This translates in a higher value of the merit function (from $\chi^2 \sim 1$ without regularisation to $\chi^2 \sim 500$ with $\lambda = 10^3$). 

\section{Comparison with usual indicators}\label{Ap:Comp}
\begin{figure}
\centering
\includegraphics[width=0.85\linewidth]{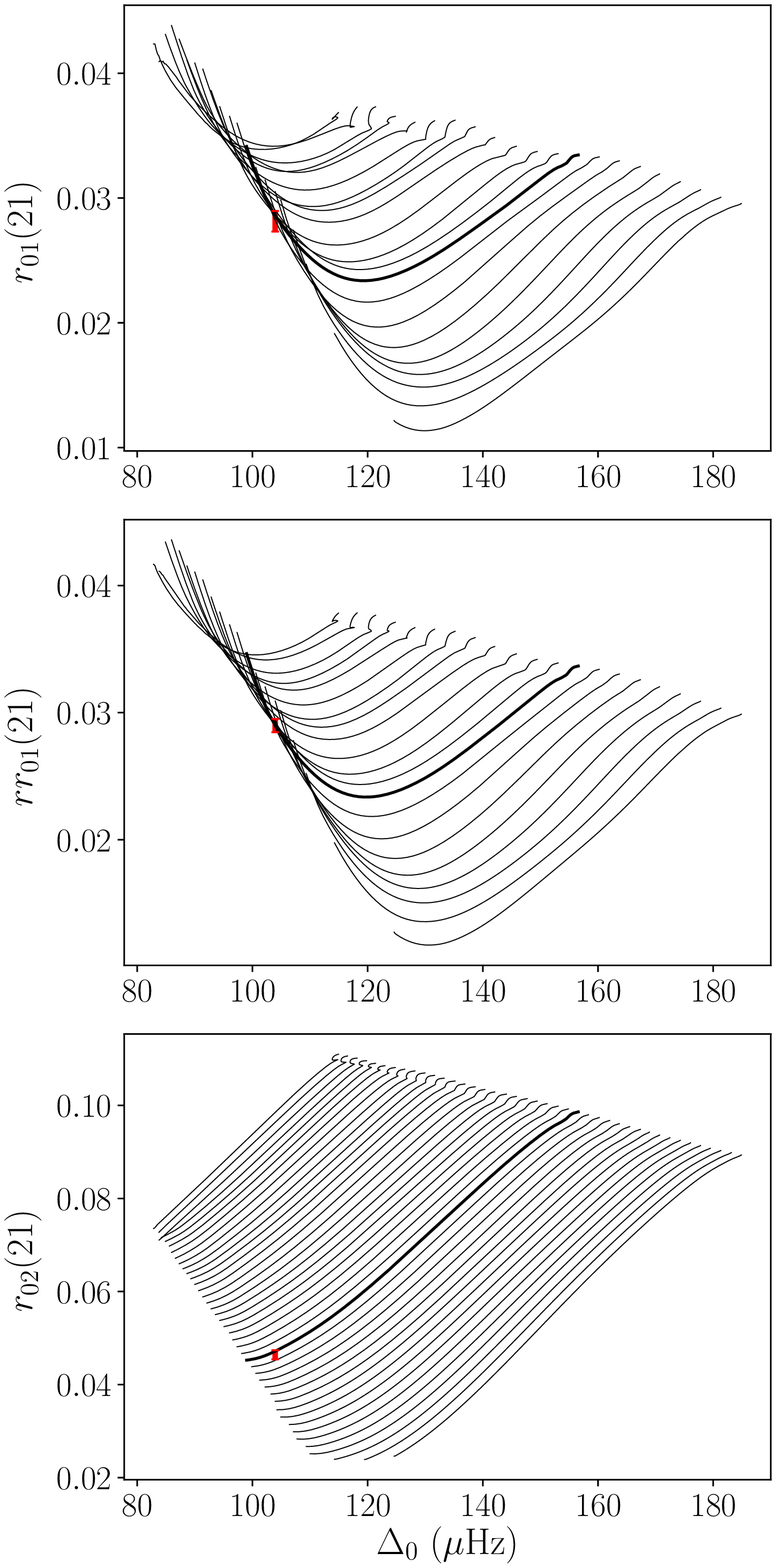}
\caption{Evolution of the normalised small separation as defined by \citet{2003A&A...411..215R} and evaluated at $n = 21$ along the grid presented in Sect.~\ref{Sec:Mod}. The red marker shows the observed value corrected for the surface effects following \citet{2008ApJ...683L.175K}'s prescription. The thick line represents the track for $1.06M_{\odot}$. The top panel is the three points normalised small separation between spherical degrees $0$ and $1$, the middle one is the five points normalised small separation between spherical degrees $0$ and $1$ and the bottom panel is the normalised small separation between spherical degrees $0$ and $2$.}\label{Fig:r0lRox}
\end{figure}

We present in Fig.~\ref{Fig:r0lRox} the evolution of the `usual' indicators $r_{01}(n)$ and $r_{02}(n)$ \citep{2003A&A...411..215R} used in asteroseismology in order to compare them with the new indicators. They are evaluated at the value of $n = 21$ which corresponds to the measured value of $n_\textrm{max}$ for 16 Cyg A ($n_\textrm{max}$ being the value of $n$ at $l=0$ of closest frequency to the $\nu_\textrm{max}$ value).
First let us draw the attention to the fact that the usual indicators hold a local value, as Eqs.~(\ref{Eq:d01}) and (\ref{Eq:d02}) show, while the new indicators are averaged over the complete set of available modes. It results from this averaging a lower standard deviation.
Then, we notice that the behaviour of both the usual indicators and the new ones follow similar trends. This means that the definition is consistent with what has been done up to now and that it should hold the same information. Therefore, it should be able to provide similar diagnostics, but with higher precisions.

\section{$\Gamma_1$ toy model}\label{Ap:Toy}
To build a toy model for $\Gamma_1$ that replicates at best its behaviour in the helium second ionisation zone we use the following hypotheses:
\begin{itemize}
\item In the helium second ionisation zone, hydrogen is fully ionised;
\item Metals are in their atomic form;
\item We consider a perfect gas.
\end{itemize}

We then define:
\begin{itemize}
\item The once ionised helium number of particles per unit volume $He^+$ and twice ionised helium number of particles per unit volume $He^{++}$ such that:
\begin{equation}
He^+ + He^{++} = \frac{Y \rho}{4 m_u},
\end{equation} 
where $m_u$ is the atomic mass unit;
\item The helium ionisation fraction: 
\begin{equation}
x_\textrm{He}~=~\frac{He^{++}}{He^++He^{++}},
\end{equation}
which equals $0$ if the helium is ionised only once and $1$ if it is fully ionised;
\item The electron number of particles per unit volume: 
\begin{equation}
e~=~\left( X + (1+x_\textrm{He})\frac{Y}{4}\right)\frac{\rho}{m_u};
\end{equation}
\item The total number of particles:
\begin{equation}
n~=~\left(2X + (2+x_\textrm{He})\frac{Y}{4}+\sum\limits_i\frac{Z_i}{A_i}\right)\frac{\rho}{m_u},
\end{equation}
where $Z_i$ is the mass fraction of the metal labeled $i$ and $A_i$ its mass number.
\end{itemize}

Using Saha's equation \citep{doi:10.1080/14786441008636148}:
\begin{equation}
Sa~=~\frac{He^{++} e}{He^+}~=~\frac{g}{h^3}\left(2\pi m_e k_B T\right)^\frac{3}{2} e^{-\frac{\chi}{k_B T}},
\end{equation}
where $g$ is the statistical weight of helium at its fundamental state, $h$ is Planck's constant, $k_B$ is Boltzmann's constant, $m_e$ the electron mass, and $\chi$ the helium second ionisation energy, we may obtain a second order equation for $x_\textrm{He}$:
\begin{equation}
\frac{Y \rho}{4 m_u}x_\textrm{He}^2 + \left[\left(X+\frac{Y}{4}\right)\frac{\rho}{m_u}+Sa\right]x_\textrm{He} - Sa~=~0,
\end{equation}
which we solve to obtain the evolution of $x_\textrm{He}$ with temperature and density. We may also derive this expression with respect to temperature at constant density. This yields:
\begin{equation}
\left.\frac{\partial x_\textrm{He}}{\partial T}\right|_\rho~=~\frac{(1-x_\textrm{He})\left(\frac{3}{2}+\frac{\chi}{k_B T}\right)\frac{m_u Sa}{\rho T}}{X+\left(\frac{1}{2}+x_\textrm{He}\right)\frac{Y}{2}+\frac{m_u}{\rho}Sa}.
\end{equation}

Then we have the following expressions:
\begin{equation}
P_T~=~\left.\frac{\partial ln P}{\partial ln T}\right|_s~=~1+\frac{eD}{n}\left(\frac{3}{2}+\frac{\chi}{k_B T}\right),
\end{equation}
\begin{equation}
\begin{aligned}
P_\rho &=~\left.\frac{\partial ln P}{\partial ln \rho}\right|_s~=~\left[\vphantom{\frac{Z_i}{A_i}}(2-D)X \right. \\ 
       &\left.+~(2+x_\textrm{He}-(1+x_\textrm{He})D)\frac{Y}{4}+\sum\limits_i\frac{Z_i}{A_i}\right]\frac{\rho}{m_u n},
\end{aligned}
\end{equation}
\begin{equation}
c_v~=~\frac{3}{2}\frac{k_B P_T}{\mu m_u}  + \frac{Y\chi}{4 m_u}\left.\frac{\partial x_\textrm{He}}{\partial T}\right|_\rho,
\end{equation}
where $\mu$ is the mean molecular weight and:
\begin{equation}
D~=~\left(\frac{e}{x_\textrm{He}(1-x_\textrm{He})}\frac{4 m_u}{Y\rho}+1\right)^{-1}.
\end{equation}

Finally, we may insert their values in the relation linking $\Gamma_1$, $P_T$, $P_\rho$, and $c_v$:
\begin{equation}
\Gamma_1~=~P_\rho + P^2_T \frac{P}{c_v \rho T}.
\end{equation}

\section{Gram-Schmidt's process and QR decomposition}\label{Ap:MatMet}
As a reminder, Gram-Schmidt's algorithm consists in the construction of orthonormal basis elements from a set of non-orthonormal basis elements.
Let us consider the element of index $j_0$ : $\boldsymbol{p}_{j_0}$. Let us also assume that we have already built the set of orthonormal basis functions up to index $j_0-1$, that is the set $\left(\boldsymbol{q}_1, \cdots , \boldsymbol{q}_{j_0-1} \right)$.
To build element $\boldsymbol{q}_{j_0}$, we first subtract to $\boldsymbol{p}_{j_0}$ its projection over the successive previous basis elements. We thus have:
\begin{equation}
\boldsymbol{u}_{j_0}~=~\boldsymbol{p}_{j_0} - \sum\limits^{j_0-1}_{j = 1} \left\langle \boldsymbol{p}_{j_0} \vert \boldsymbol{q}_{j} \right\rangle \boldsymbol{q}_{j},
\end{equation}
where $\boldsymbol{u}_{j_0}$ is the basis element orthogonal to the set $\left(\boldsymbol{q}_1, \cdots , \boldsymbol{q}_{j_0-1} \right)$. Finally, it is normalised to obtain:
\begin{equation}
\boldsymbol{q}_{j_0}~=~\frac{\boldsymbol{u}_{j_0}}{\Vert \boldsymbol{u}_{j_0} \Vert}.
\end{equation}

It is also possible to express this process as a QR decomposition. To do so we call $\mathcal{P}_l$ the matrix of initial polynomials and $\mathcal{Q}_l$ the matrix of orthonormal polynomials for a given spherical degree:
\begin{equation}
\mathcal{P}_l~=~
\begin{pmatrix}
   p_0(n_\textrm{min}) & \cdots & p_{2}(n_\textrm{min}) \\
   \vdots &  &  \vdots \\
   p_0(n_\textrm{max}) & \cdots & p_{2}(n_\textrm{max}) \\
\end{pmatrix}
~=~
\begin{pmatrix}
	\boldsymbol{p}_0 & \cdots & \boldsymbol{p}_{2}
\end{pmatrix}_l,
\end{equation}
\begin{equation}
\mathcal{Q}_l~=~
\begin{pmatrix}
   q_0(n_\textrm{min}) & \cdots & q_{2}(n_\textrm{min}) \\
   \vdots &  &  \vdots \\
   q_0(n_\textrm{max}) & \cdots & q_{2}(n_\textrm{max}) \\
\end{pmatrix}
~=~
\begin{pmatrix}
	\boldsymbol{q}_0 & \cdots & \boldsymbol{q}_{2}
\end{pmatrix}_l,
\end{equation}

where $n_\textrm{min}$ is the lowest observed radial order and $n_\textrm{max}$ the highest one for the spherical degree considered.

We may then express \emph{Gram-Schmidt}'s procedure in a matrix form as a QR decomposition as follows:
\begin{equation}
\mathcal{Q}_l~=~\mathcal{P}_lR^{-1}_l,
\end{equation}
where, using our definition of the scalar product:
\begin{equation}
R_l~=~
\begin{pmatrix}
   \left\langle \boldsymbol{q}_0 \vert \boldsymbol{p}_{0} \right\rangle_l & \cdots & \left\langle \boldsymbol{q}_0 \vert \boldsymbol{p}_{2} \right\rangle_l \\
    & \ddots &  \vdots \\
  0 & & \left\langle \boldsymbol{q}_{2} \vert \boldsymbol{p}_{2} \right\rangle_l \\
\end{pmatrix},
\end{equation}
is an upper triangular matrix.

We may now generalise by considering several spherical degrees. For the smooth part, the matrices become block matrices which are the combination of the matrices for the several spherical degrees considered. This block disposition illustrates the fact that we have different basis functions for the different spherical degrees -- this is not the case for the glitch part --. This is represented by a \emph{Kronecker} delta in Eq. \ref{Eq:PolSmo}. In these matrices, each row corresponds to a given mode and each column corresponds to a given element of the basis.
Finally, we append the glitch part to those block matrices. They take the following form (we remind that the independent variable for the glitch basis is $\widetilde{n}~=~n + l/2$): 

\scriptsize{
\begin{equation}
\newcommand*{\temp}[1]{\multicolumn{1}{|c}{#1}}
\mathbb{P}~=~
\left(
\begin{array}{cccccc}
   \mathcal{P}_0 & & 0 & \temp{\boldsymbol{p}_\textrm{He,C,5}(n)} & \cdots & \boldsymbol{p}_\textrm{C,S}(n) \\
    & \ddots & & \temp{\vdots} & & \vdots \\
  0 &  & \mathcal{P}_3 & \temp{\boldsymbol{p}_\textrm{He,C,5}(n+3/2)} & \cdots & \boldsymbol{p}_\textrm{C,S}(n+3/2) \\
\end{array}\right),
\end{equation}
\begin{equation}
\newcommand*{\temp}[1]{\multicolumn{1}{|c}{#1}}
\mathbb{Q}~=~
\left(
\begin{array}{cccccc}
   \mathcal{Q}_0 & & 0 & \temp{\boldsymbol{q}_\textrm{He,C,5}(n,l)} & \cdots & \boldsymbol{q}_\textrm{C,S}(n,l) \\
    & \ddots & & \temp{\vdots} & & \vdots \\
  0 &  & \mathcal{Q}_3 & \temp{\boldsymbol{q}_\textrm{He,C,5}(n,l)} & \cdots & \boldsymbol{q}_\textrm{C,S}(n,l) \\ 
\end{array}\right),
\end{equation}
}
\normalsize
And the $R$ matrix becomes:

\scriptsize{
\begin{equation}
\newcommand*{\temp}[1]{\multicolumn{1}{|c}{#1}}
\mathbb{R}~=~
\left(
\begin{array}{cccccc}
R_0 & & 0 & \temp{\overrightarrow{\left\langle \boldsymbol{q}_k \vert \boldsymbol{p}_\textrm{He,C,5} \right\rangle}_{l=0}} & \cdots & \overrightarrow{\left\langle \boldsymbol{q}_k \vert \boldsymbol{p}_\textrm{C,S} \right\rangle}_{l=0} \\
& \ddots & & \temp{\vdots} & & \vdots \\
0 & & R_3 & \temp{\overrightarrow{\left\langle \boldsymbol{q}_k \vert \boldsymbol{p}_\textrm{He,C,5} \right\rangle}_{l=3}} & \cdots & \overrightarrow{\left\langle \boldsymbol{q}_k \vert \boldsymbol{p}_\textrm{C,S} \right\rangle}_{l=3} \\ \cline{1-3}
\multicolumn{3}{c}{\multirow{3}{*}{0}} & \left\langle \boldsymbol{q}_\textrm{He,C,5} \vert \boldsymbol{p}_\textrm{He,C,5} \right\rangle & \cdots & \left\langle \boldsymbol{q}_\textrm{He,C,5} \vert \boldsymbol{p}_\textrm{C,S} \right\rangle \\
\multicolumn{3}{c}{} & & \ddots & \vdots \\
\multicolumn{3}{c}{} & 0 & & \left\langle \boldsymbol{q}_\textrm{C,S} \vert \boldsymbol{p}_\textrm{C,S} \right\rangle\\
\end{array}\right),
\end{equation}}
\normalsize
where $\overrightarrow{\left\langle \boldsymbol{q}_k \vert \boldsymbol{p}_g \right\rangle}_{l=0}$ is a column vector whose rows are the successive scalar products of the basis elements with the glitch function -- denoted by the $g$ index -- such that $\overrightarrow{\left\langle \boldsymbol{q}_k \vert \boldsymbol{p}_g \right\rangle}_{l=0} = \begin{pmatrix}
\left\langle \boldsymbol{q}_0 \vert \boldsymbol{p}_g \right\rangle_{l=0} & \cdots & \left\langle \boldsymbol{q}_{2} \vert \boldsymbol{p}_g \right\rangle_{l=0}
\end{pmatrix}^T$. We may again write the QR decomposition as:
\begin{equation}\label{Eq:QR}
\mathbb{Q}~=~\mathbb{P} \mathbb{R}^{-1}
\end{equation}

\section{A numerical example}\label{Ap:NEx}
\begin{figure}
\centering
\includegraphics[width=0.85\linewidth]{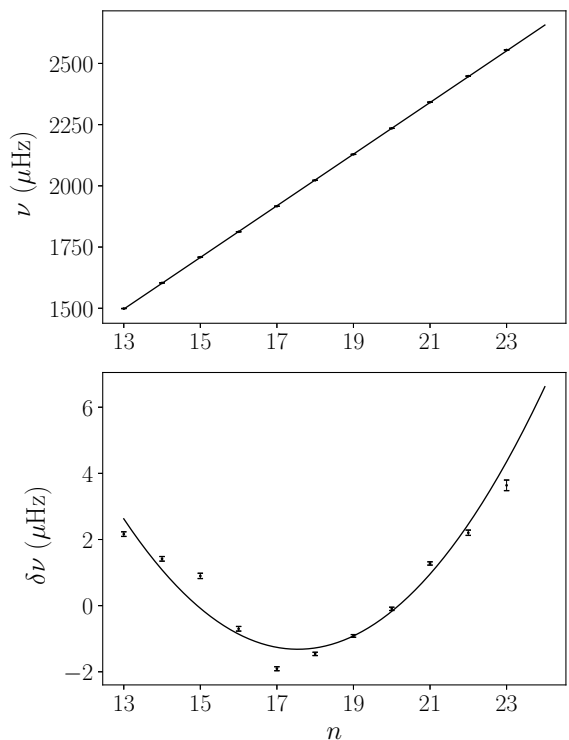}
\caption{Comparison between the successive adjustments and the observed radial modes frequencies listed in Table \ref{Tab:Freq}. The upper panel represents the observed frequencies compared to the first order adjustment while the lower panel is the residual of the first order adjustment compared to the second order fit to those residuals.}\label{Fig:Gra}
\end{figure}

\begin{figure}
\centering
\includegraphics[width=0.85\linewidth]{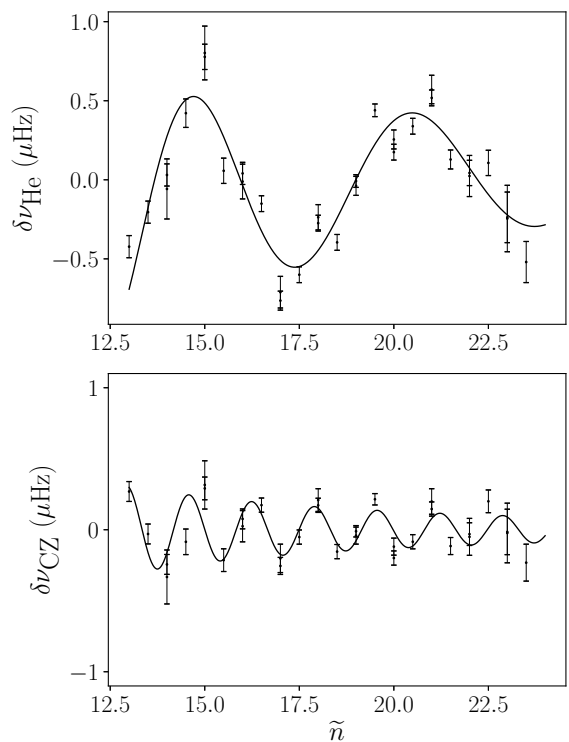}
\caption{Separated glitches adjustments to the frequencies in Table \ref{Tab:Freq} of both glithes shown for the spherical degree $l=0$. The upper panel shows the helium glitch while the lower one shows the convection zone glitch.}\label{Fig:GraGli}
\end{figure}

We generated a set of frequencies for spherical degrees from $0$ to $2$ (listed in Table \ref{Tab:Freq}) and applied our method. We show in Figs. \ref{Fig:Gra} and \ref{Fig:GraGli} the successive adjustments of the basis functions. This shows the validity of using the set of functions described in Sect.~\ref{Sec:Met} and allows oneself to compare their results with ours. As a reminder, we project the frequencies over the basis in the specific order explained in Sect. \ref{Sec:Met}. Therefore, taken in the correct order, these plots provide intermediary results for the glitch adjustment. We also show in Table~\ref{Tab:alk} the values of the fitted coefficients. In this example, we also fitted $n^{-1}$ polynomials  -- inspired by the second order form of the asymptotic expansion -- to show that it is not necessary, nor relevant, to add supplementary basis elements to our method. Indeed, the fitted coefficient values become comparable to the standard deviation that is equal to $1$, through the orthonormalisation. Such values and plots were obtained as follows:
\begin{enumerate}
\item Considering the set of standard deviations from Table~\ref{Tab:Freq}, we use Gram-Schmidt procedure (Eq.~\ref{Eq:Bas}) associated with the definition of the scalar product (Eq.~\ref{Eq:ScaPro}) to produce the ordered orthonormal basis functions for each value of the spherical degree $l$. Thus, we successively project the former basis elements $p_j(n,l)$ (i.e. the ordered set of polynomials $n^0$,$n^1$,$n^2$ and the glitch functions for each spherical degree) on the already defined orthonormal basis elements $q_{j_0}(n,l)$. Then, we normalise those projections. This provides us with the orthonormal basis elements $q_{j_0}(n,l)$. as well as the transformation matrix $R^{-1}_{j,j_0}$.
\item For the smooth part and one spherical degree at a time, the frequencies from Table~\ref{Tab:Freq} are projected on the orthonormal basis elements following the proper order to produce the fitted frequencies, $\nu_f(n,l)$, according to Eq.~\ref{Eq:FreFit}. Where the fitted coefficients are given by Eq.~\ref{Eq:aSmo}. Then, for the glitch part, the frequencies are projected simultaneously for every spherical degree on the glitch basis elements following the same procedure. This is due to the fact that the glitch coefficients should not depend on $l$. This produces the coefficients from Table~\ref{Tab:alk}.
\end{enumerate}

\begin{table}
\centering
\begin{tabular}{cccc}
\hline
$l$ & $n$ & $\nu (\mu Hz)$ & $\sigma (\mu Hz)$ \\[0.2em]
\hline
\hline\\[-0.8em]
$0$ & $13$ & $1498.89$ & $0.07$ \\
$0$ & $14$ & $1603.60$ & $0.07$ \\
$0$ & $15$ & $1708.55$ & $0.08$ \\
$0$ & $16$ & $1812.40$ & $0.07$ \\
$0$ & $17$ & $1916.65$ & $0.06$ \\
$0$ & $18$ & $2022.56$ & $0.05$ \\
$0$ & $19$ & $2128.56$ & $0.04$ \\
$0$ & $20$ & $2234.84$ & $0.05$ \\
$0$ & $21$ & $2341.67$ & $0.05$ \\
$0$ & $22$ & $2448.06$ & $0.08$ \\
$0$ & $23$ & $2554.95$ & $0.16$ \\
$1$ & $13$ & $1546.42$ & $0.07$ \\
$1$ & $14$ & $1651.36$ & $0.09$ \\
$1$ & $15$ & $1755.56$ & $0.08$ \\
$1$ & $16$ & $1860.40$ & $0.05$ \\ 
$1$ & $17$ & $1965.44$ & $0.05$ \\
$1$ & $18$ & $2071.47$ & $0.05$ \\
$1$ & $19$ & $2178.50$ & $0.04$ \\
$1$ & $20$ & $2284.98$ & $0.05$ \\
$1$ & $21$ & $2391.77$ & $0.06$ \\
$1$ & $22$ & $2499.11$ & $0.08$ \\
$1$ & $23$ & $2606.15$ & $0.13$ \\
$2$ & $13$ & $1596.42$ & $0.19$ \\
$2$ & $14$ & $1701.68$ & $0.17$ \\
$2$ & $15$ & $1805.69$ & $0.11$ \\
$2$ & $16$ & $1910.30$ & $0.10$ \\
$2$ & $17$ & $2016.47$ & $0.08$ \\
$2$ & $18$ & $2122.70$ & $0.06$ \\
$2$ & $19$ & $2229.41$ & $0.06$ \\
$2$ & $20$ & $2336.56$ & $0.09$ \\
$2$ & $21$ & $2443.24$ & $0.13$ \\
$2$ & $22$ & $2550.54$ & $0.21$ \\[0.2em]
\hline
\end{tabular}
\caption{Example set of frequencies.}\label{Tab:Freq}
\end{table}

\begin{table}
\centering
\begin{tabular}{cccc}
\hline
 & $l=0$ & $l=1$ & $l=2$ \\[0.2em]
\hline
\hline\\[-0.8em]
 $a_{l0}$ & $112948.0$ & $117296.1$ & $72112.2$ \\
 $a_{l1}$ & $14952.0$ & $14674.1$ & $6882.0$ \\
 $a_{l2}$ & $74.0$ & $79.4$ & $35.4$ \\
 $a_{l-1}$ & $-3.5$ & $9.0$ & $6.3$ \\[0.2em]
 \hline
 \hline\\[-0.8em]
 $A_\textrm{He}$ & \multicolumn{3}{c}{$27.5$} \\
 $A_\textrm{CZ}$ & \multicolumn{3}{c}{$9.4$} \\[0.2em]
\hline
\end{tabular}
\caption{Fitted parameters to the frequencies of Table \ref{Tab:Freq}.}\label{Tab:alk}
\end{table}

\section{Supplementary adjustments}\label{Ap:SupAdj}
\begin{figure}
\centering
\includegraphics[width=0.85\linewidth]{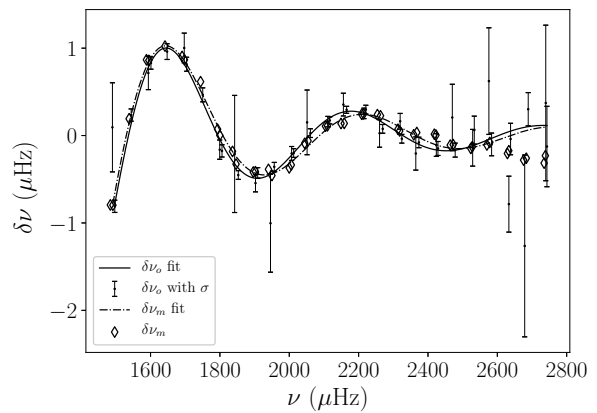}
\caption{Comparison between the observed helium glitch $\delta\nu_o$ (solid line) and the one resulting from the best fit model $\delta\nu_m$ (dot-dashed line) for $l=0$. We also display the observed glitch as a function of the frequencies (errorbars) as well as the best model glitch associated with the theoretical frequencies (diamond). The observed frequencies have been corrected for surface effects using \citet{2015A&A...583A.112S}'s prescription.}\label{Fig:GliSon}
\end{figure}

\begin{figure}
\centering
\includegraphics[width=0.85\linewidth]{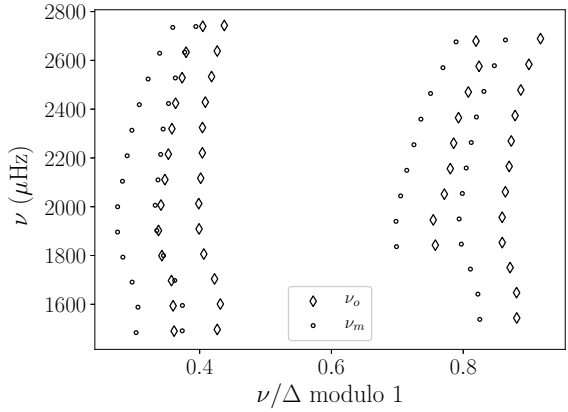}
\caption{Comparison between the observed frequencies (diamonds) and the best model frequencies (circles) in an \emph{échelle} diagram. The observed frequencies have been corrected for surface effects using \citet{2015A&A...583A.112S}'s prescription.}\label{Fig:EchSon}
\end{figure}

We show in the present section the best fit model frequencies adjusted to the observed 16 Cygni A frequencies for several cases.
First we show the adjustment to the frequencies corrected for surface effects using a Lorentzian profile as in \citet{2015A&A...583A.112S}. Fig. \ref{Fig:GliSon} shows that both glitches are in good agreement. However, the frequencies are systematically shifted, as Fig. \ref{Fig:EchSon} illustrates, as a consequence of the difference between the observed and theoretical $\hat{\epsilon}$ values. Fig. \ref{Fig:EpsEvo} illustrates such a discrepancy. This shows that, even though \citet{2015A&A...583A.112S} showed that for high frequency regimes \citet{2008ApJ...683L.175K}'s prescription is not able to reproduce frequency differences between patched and unpatched model, it is the only tested empirical correction that allowed us to reproduce the observed value for $\hat{\epsilon}$. In a further study, it would be appropriate to try out a scaled formulation of \citet{2014A&A...568A.123B}'s correction such as presented in \citet{2018arXiv180908904M}.

\begin{figure}
\centering
\includegraphics[width=0.85\linewidth]{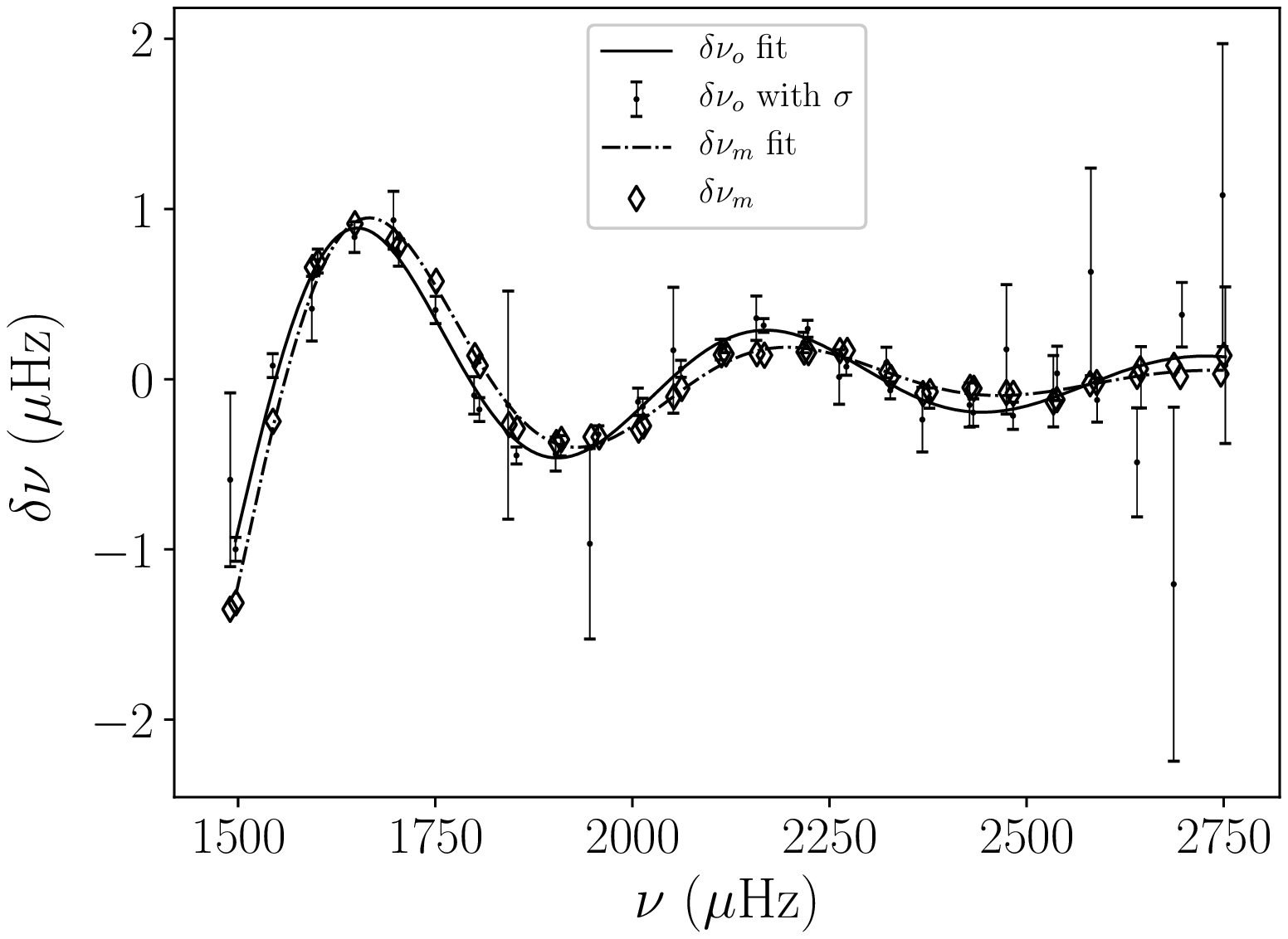}
\caption{Comparison between the observed helium glitch $\delta\nu_o$ (solid line) and the one resulting from the best fit model $\delta\nu_m$ (dot-dashed line) for $l=0$. We also display the observed glitch as a function of the frequencies (errorbars) as well as the best model glitch associated with the theoretical frequencies (diamond). We include polynomials in $n^{-1}$ to the basis functions.}\label{Fig:Gli-1}
\end{figure}

We also provide in Fig. \ref{Fig:Gli-1} the best fit model including terms in $n^{-1}$ in the basis functions for the smooth part of the spectrum. By eye, the glitch adjustment seems better than in Fig. \ref{Fig:ObsFit}. However, as shown in table \ref{Tab:alk}, the improvement is not significant as the fitted parameters values are comparable to their standard deviation. Moreover, Table \ref{Tab:StePar} demonstrates that the effect of including such terms in the adjustment is comparable to a variation of $1 \sigma$ in the frequencies.

\end{appendix}

\end{document}